\documentclass[colorlinks=true,linkcolor=black,citecolor=blue,urlcolor=blue]{aa}  

\usepackage[version=4]{mhchem}
\usepackage{multirow}
\usepackage{ulem}
\usepackage{graphicx}
\usepackage{xcolor}
\usepackage{orcidlink}
\usepackage{amsmath}
\usepackage{txfonts}
\definecolor{dblue}{rgb}{0.0, 0.0, 0.55}

\def\la{\mathrel{\mathchoice {\vcenter{\offinterlineskip\halign{\hfil
$\displaystyle##$\hfil\cr<\cr\sim\cr}}}
{\vcenter{\offinterlineskip\halign{\hfil$\textstyle##$\hfil\cr
<\cr\sim\cr}}}
{\vcenter{\offinterlineskip\halign{\hfil$\scriptstyle##$\hfil\cr
<\cr\sim\cr}}}
{\vcenter{\offinterlineskip\halign{\hfil$\scriptscriptstyle##$\hfil\cr
<\cr\sim\cr}}}}}

  \def\st{{\circ\hspace{-0.86ex}-}}
  \def\pst{{p^\st}}
  \def\Tst{{T^\st}}
  \def\ust{u^\st}
  \def\sst{s^\st}
  \def\must{{\mu^\st}}
  \def\neqq{{\stackrel{\circ}{n}}}
  \def\peq{{\stackrel{\circ}{p}}}
  \def\Neq{{\stackrel{\circ}{N}}}
  \def\dH{{\Delta_{\rm f}H^{\hspace{0.1ex}\st}\!}}
  \def\dHR{{\Delta_{\rm R}H^{\hspace{0.1ex}\st}\!}}
  \def\dG{{\Delta_{\rm f}G^{\hspace{0.1ex}\st}\!}}
  \def\dGR{{\Delta_{\rm R}G^{\hspace{0.1ex}\st}\!}}
  \def\nH{n_{\langle\rm H\rangle}}

\begin{document} 

\title {Can thermodynamic equilibrium be established in planet-forming disks?}

\author{Jayatee Kanwar \orcidlink{0000-0003-0386-2178} \inst{1,2,3} 
        \and
        Peter Woitke \orcidlink{0000-0002-8900-3667} \inst{2}
        \and
        Inga Kamp \orcidlink{0000-0001-7455-5349} \inst{1}
        \and
        Paul Rimmer\inst{4}
        \and
        Christiane Helling\inst{2}
}

\institute{Kapteyn Astronomical Institute, University of Groningen, PO Box 800, 9700 AV Groningen, The Netherlands
\and
Space Research Institute, Austrian Academy of Sciences, Schmiedlstr.~6, A-8042, Graz, Austria
\and
Institute for Theoretical Physics and Computational Physics, Graz University of Technology, Petersgasse 16, 8010 Graz, Austria
\and
Cavendish Laboratory, University of Cambridge, JJ Thomson Ave, Cambridge, CB3 0HE, United Kingdom           
}

  \abstract
   {The inner regions of planet-forming disks are warm and dense. The chemical networks used for disk modelling so far were developed for a cold and dilute medium and do not include a complete set of pressure-dependent reactions. The chemical networks developed for planetary atmospheres include such reactions along with the inverse reactions related to the Gibb's free energies of the molecules. The chemical networks used for disk modelling are thus incomplete in this respect.}
   {We want to study whether thermodynamic equilibrium can be established in a planet-forming disk. We identify the regions in the disk most likely to reach thermodynamic equilibrium and determine the timescale over which this occurs.}
   {We employ the theoretical concepts used in exoplanet atmosphere chemistry for the disk modelling with  PROtoplanetary DIsk MOdel ({\sc ProDiMo}). We develop a chemical network called CHemistry Assembled from exoplanets and dIsks for Thermodynamic EquilibriA ({\sc ChaiTea}) that is based on the UMIST 2022, STAND, and large DIscANAlysis (DIANA) chemical networks. It consists of 239 species. From the STAND network, we adopt the concept of reversing all gas-phase reactions based on thermodynamic data. We use single-point models for a range of gas densities and gas temperatures to verify that the implemented concepts work and thermodynamic equilibrium is achieved in the absence of cosmic rays and photoreactions including radiative associations and direct recombinations. We then study the impact of photoreactions and cosmic rays that lead to deviations from thermodynamic equilibrium. We explore the chemical relaxation timescales towards thermodynamic equilibrium. Lastly, we study the predicted 2D chemical structure of a typical T\,Tauri disk when using the new {\sc ChaiTea} network instead of the large DIANA standard network, including photorates, cosmic rays, X-rays, and ice formation.}
   {We find that abundances calculated with {\sc ProDiMo} using the {\sc ChaiTea} network agree with those from the equilibrium chemistry code Gleich-Gewichts-Chemie (GGchem) down to 600\,K when the photorates, cosmic rays, and X-rays (benchmark) are absent. To measure the deviation between thermodynamic equilibrium and chemical kinetics, a measure, $\sigma$, is introduced that evaluates the mean logarithmic deviation between the two abundance sets, which is $<1\%$ in the benchmark case. In the presence of photoreactions, based on a local Planck radiation field, $\sigma$ increases to $\sim\!0.1$ across all densities and temperatures. When the cosmic-ray ionisation rate is increased from zero to about $10^{-25}\rm\,s^{-1}$, $\sigma$ begins to become large ($>\!1$), affecting in particular the ions, and when it reaches the standard value of $10^{-17}\rm\,s^{-1}$, $\sigma$ becomes $>$\,10. Low-temperature and low-density regions are more affected than high-temperature and high-density regions, as expected. The chemical relaxation timescales show a wide range, with both slow and fast chemical processes. The 2D disk models show that thermodynamic equilibrium cannot be established anywhere in the disk. However, a tiny, warm, high-density region directly behind the inner rim which is shielded from the cosmic rays, approaches thermodynamic equilibrium to some extent. In the warm intermittent molecular layer, which is observable, $\sigma\,\geq\,10$, and $\sigma$ is even higher in other regions.}
   {We have developed a new chemical network, {\sc ChaiTea}, that merges planetary chemistry with disk chemistry and have implemented it in the 2D thermochemical disk code {\sc ProDiMo}. The inclusion of termolecular and reverse reactions changes the chemical structure in the inner disk that is probed by JWST.}
  \maketitle
%

\section{Introduction}

Astrochemical models for interstellar clouds and protoplanetary disks differ in various ways from models simulating the chemistry in planetary and exoplanetary atmospheres.  Using the kinetic rate networks originally developed for interstellar clouds makes sense for relatively cold and tenuous outer disks. However, the densities and temperatures in the inner disks around T\,Tauri stars resemble the conditions in planetary atmospheres where the equilibrium chemistry approach is often used \citep{Marley2015}. This raises the questions of where and how these astrochemical models and methods developed separately for interstellar clouds and disks, and planetary atmospheres connect in terms of density and temperatures. 

Studies such as \cite{Lodders2002} and \cite{Fortney2006} discussed the departure from thermodynamic equilibrium in planetary atmospheres of hot Jupiters and demonstrated that disequilibrium chemistry better explains the observed spectra. \cite{Cooper2006} studied the effect of disequilibrium chemistry and reported that chemical equilibrium may not exist in the exoplanetary atmosphere due to long timescales at low temperatures and pressure. \cite{Carleo2022} and \cite{Guilluy2022} used thermodynamic equilibrium to analyse the observations of warm Saturn-mass planets but suspected that disequilibrium chemistry can better interpret them.  \cite{Zahnle2009b}, \cite{Zahnle2009}, \cite{Moses2011}, \cite{Venot2014}, and \cite{Barth2021} studied (photo-) chemical kinetics. \cite{Tsai2023} used photochemically produced \ce{SO2} to explain the observed feature at 4.05\,$\mu$m in the transmission spectrum of WASP-39b, \cite{Molaverdikhani2020} demonstrate the effect of quenching in ultra-hot Jupiters, and \cite{Moses2011} and \cite{Moses2014} reported that thermodynamic equilibrium can only prevail in deeper layers of hot planetary atmospheres. \cite{Blumenthal2018} studied whether the differences caused by equilibrium and disequilibrium chemistry are observable across the range of exoplanets and report that disequilibrium chemistry should be considered to determine the metallicity of a planet. \cite{Lodders2003} used thermodynamic equilibrium approach to calculate condensation temperatures for a wide range of gas-phase species and condensates.

One such models developed to simulate the chemistry in planetary and exoplanetary atmospheres is ARGO \citep{Rimmer2016, Rimmererr2019}, which introduced the STAND chemical rate network. This network includes a large number of three-body reactions, it has reaction-rate coefficients in the low- and the high-pressure limits, and it uses a thermodynamical concept to reverse reactions based on the temperature-dependent Gibbs free energies of the molecules. This concept is key to ensuring that, at very high pressures, the abundances approach their thermodynamic equilibrium values. 

In this paper we aim to utilise these concepts and implement them in the radiative thermochemical disk model PROtoplanetary DIsk MOdel ({\sc ProDiMo}) \citep{Woitke2009, Kamp2017, Woitke2016}. We merged this exoplanetary chemical network with the chemical network developed for disks and aimed to test the validity of thermodynamic equilibrium in the planet-forming disk environments. Section\,\ref{Theoretical chemistry} explains the theoretical concepts adopted from the exoplanet community. Section\,\ref{Implementation} describes our methodology and the modelling results are shown in Sect.\,\ref{Results}. The effect of reversing the reactions on the chemical abundances and analysing whether thermodynamic equilibrium is established in disk models are studied in Sect.\,\ref{Discussion_ch4}. We present our conclusion in Sect.\,\ref{Conclusion_ch4}.

\section{Chemical reactions}
\label{Theoretical chemistry}

Here we briefly summarise the theoretical concepts used to determine the rate coefficients for the different types of chemical reactions. \cite{Kamp2017} included photoreactions, a few three-body reactions and chemical concepts from the interstellar medium. We used the photo-cross-sections for the photoreactions from the Leiden database \citep{Dishoeck2006, Heays2017}. The bimolecular rates were calculated using the modified Arrhenius equation described by \cite{Arrhenius1889}. Below, we explain the concepts introduced now from exoplanet chemistry models. 

\subsection{Lindemann-Hinshelwood mechanism}
\label{3body}

There are a few reaction types that are regularly included in planetary chemistry, but are not typically considered in interstellar and disk chemistry because they are irrelevant for the low pressures and temperatures in interstellar clouds. These reactions proceed via an activated complex that is stabilised by another collision with an abundant third gas particle, for example    
\begin{equation*}
  \hspace*{-1mm}\begin{array}{lrcl}
    \mbox{(1) termolecular} & \rm A ~+~ B ~+~ M &\to& \rm AB ~+~ M \\[1mm]
    \mbox{(2) thermal decomp.} & \rm AB ~+~ M &\to& \rm A ~+~ B ~+~ M \\[1mm]
    \mbox{(3) 3-body recomb.} & \rm A^+ ~+~ e^- ~+~ M &\to& \rm A ~+~ M \\[1mm]
    \mbox{(4) thermal ionization} & \rm A ~+~ M &\to& \rm A^+ ~+~ e^- ~+~ M 
  \end{array}
\end{equation*}
where A, B, and AB are atoms or molecules.  The (2) and (4) reaction types, which are endothermic and have large activation energies, can be considered as the reverse of the (1) and (3) reaction types, respectively. Another example is the termolecular ion-neutral reaction where A and AB are charged molecules.  

The reaction rates of these types of reactions are assumed to follow the Lindemann-Hinshelwood mechanism \citep{Lindemann1922}, which is derived in \cite{Atkin2006} as
\begin{equation} \label{Lind}
    K = \frac{K_0\,n_{\rm M}} {1+K_0\,n_{\rm M}/K_{\infty}} ,
\end{equation}
where $K_0$ and $K_{\infty}$ are the rate constants in the low- and the high-pressure limits, respectively. The unit of $K$ depends on the reaction type and the pressure regime, and $n_{\rm M}$ is the number density [cm$^{-3}$] of the third body.

\begin{eqnarray}
    K_\circ &=& \alpha_0\,\biggl({\frac{T}{300}}\biggl)^{\beta_0} \exp\Big(-\frac{\gamma_0}{T}\Big) 
    \label{Arrhenius} \\
    K_\infty &=& \alpha_\infty\,\biggl({\frac{T}{300}}\biggl)^{\beta_\infty} \exp\Big(-\frac{\gamma_\infty}{T}\Big)
\end{eqnarray}
In the case of a termolecular or three-body recombination reaction, the unit of $K_{\circ}$ is cm$^{6}$s$^{-1}$ and the units of $K_\infty$ and $K$ are cm$^{3}$s$^{-1}$. 
In {\sc ProDiMo}, we assume $n_{\rm M}=n_{\rm H_2} + n_{\rm H} + n_{\rm H^+} + n_{\rm He} + n_{\rm He^+}$ for the particle density of the third body. There are six constants required to calculate the effective reaction rate, $K$. 

In some works, the reaction rate according to Eq.\,(\ref{Lind}) is multiplied by another factor, $F$, which is a dimensionless function of temperature and pressure \citep{Troe1983}. This is used to obtain a more accurate transition from the low- to the high-pressure limit. We neglect these corrections in this work.

\subsection{Backward reactions}
\label{backward}

We consider an example of a pair of a forward and a reverse gas phase reaction
\begin{equation}
 \rm A ~+~ B ~\leftrightarrow~ C ~+~ D ~+~ E \ .
\end{equation}
In thermodynamic equilibrium (TE), the law of mass action (see Appendix~\ref{LawMassAction}) for this reaction reads 
\begin{equation}
  \left(\frac{\peq_{\rm C}}{\pst}\right) 
  \left(\frac{\peq_{\rm D}}{\pst}\right) 
  \left(\frac{\peq_{\rm E}}{\pst}\right) 
  ~=~
  \left(\frac{\peq_{\rm A}}{\pst}\right)
  \left(\frac{\peq_{\rm B}}{\pst}\right)
  ~\exp\left(-\frac{\dGR}{RT}\right) 
  \label{MassAction}
\end{equation}
where $\peq_i\!=\!\neqq_i\,kT$ is the partial pressure,  $\neqq_i$ the number density of species, $i$, in thermodynamic equilibrium, and $\pst\!\!=\!1$\,bar is a standard pressure. $\dGR\rm\,[J/mol]$ is the change in Gibbs free energy per mol of reactions
\begin{equation}
  \dGR = \dG({\rm C})+\dG({\rm D})+\dG({\rm E})-\dG({\rm A})-\dG({\rm B})
\end{equation}
and the $\dG$ are the Gibbs free energies of formation of the reactants and products as pure substances from the standard states of the elements at room temperature. In thermodynamic equilibrium, the forward and reverse reaction rates are equal (detailed balance), hence
\begin{equation}
   K_{\rm f}\;\neqq_{\rm A}\,\neqq_{\rm B}
 ~=~ K_{\rm r}\;\neqq_{\rm C}\,\neqq_{\rm D}\, \neqq_{\rm E}  \ ,
 \label{DetailedBalance}
\end{equation}
where $K_{\rm f}$ and $K_{\rm r}$ are the rate coefficients of the forward and reverse reactions.  Combining Eqs.~(\ref{MassAction}) and (\ref{DetailedBalance}) results in
\begin{equation}
  \frac{K_{\rm r}}{K_{\rm f}} 
  = \frac{\neqq_{\rm A}\,\neqq_{\rm B}}{\neqq_{\rm C}\,\neqq_{\rm D}\,\neqq_{\rm E}}
  = \biggl({\frac{\pst}{kT}}\biggl)^{N} 
    \exp\left(+\frac{\dGR}{RT}\right),
  \label{kreverse}  
\end{equation}
where $N$, in general, is the number of reactants minus the number of products; here $N=-1$.  

\begin{table*}
\caption{Elements and chemical gas-phase species in the network.}
\label{tab:species}
\resizebox{13cm}{!}{
\begin{tabular}{|c|l|c|}
\hline
12 elements & H, He, C, N, O, Ne, Na, Mg, Si, S, Ar, Fe& \\
\hline
(H) & H, H$^+$, H$^-$, {\bf H$_2$}, H$_2^+$, H$_3^+$ & 6 \\
(He) & He, He$^+$ & 2 \\
(He-H) & HeH$^+$ & 1 \\
(C-H) & C, C$^+$, C$^{++}$, CH, CH$^+$, CH$_2$, CH$_2^+$,
CH$_3$, CH$_3^+$, {\bf CH$_4$}, CH$_4^+$, CH$_5^+$ & 12 \\
(C-C) & {\bf C$_2$}, C$_2^+$, C$_2$H, C$_2$H$^+$, {\bf C$_2$H$_2$}, C$_2$H$_2^+$, C$_2$H$_3$, C$_2$H$_3^+$, {\bf C$_2$H$_4$}, C$_2$H$_4^+$, C$_2$H$_5$, C$_2$H$_5^+$, & \\
& {\bf C$_3$}, C$_3^+$, C$_3$H, C$_3$H$^+$, {\bf C$_3$H$_2$},
C$_3$H$_2^+$, C$_3$H$_3^+$, & \\
& {\bf C$_4$}, C$_4^+$, C$_4$H$^+$ & 23 \\
(C-N) & CN, CN$^+$, {\bf HCN}, HCN$^+$, HCNH$^+$, HNC, H$_2$CN, OCN, OCN$^+$ & 9 \\
(C-O) & {\bf CO}, CO$^+$, HCO, HCO$^+$, & \\
& {\bf CO$_2$}, CO$_2^+$, HCO$_2^+$, C$_2$O, C$_2$O$^+$, HC$_2$O$^+$, & \\
& {\bf H$_2$CO}, H$_2$CO$^+$, CH$_3$O, H$_3$CO$^+$, CH$_2$OH, & \\
& {\bf CH$_3$OH}, CH$_3$OH$^+$, CH$_3$OH$_2^+$ & 18 \\
(C-S) & {\bf CS}, CS$^+$, HCS, HCS$^+$, {\bf H$_2$CS}, H$_2$CS$^+$, H$_3$CS$^+$, & \\
& {\bf OCS}, OCS$^+$, HOCS$^+$ & 10 \\
(N-H) & N, N$^+$, N$^{++}$, NH, NH$^+$, NH$_2$, NH$_2^+$, {\bf NH$_3$},
NH$_3^+$, NH$_4^+$ & 10 \\
(N-N) & {\bf N$_2$}, N$_2^+$, HN$_2^+$, \textbf{\ce{N2O}}, \ce{N2O+}, \ce{HN2O+}& 6 \\
(N-O) & NO, NO$^+$, {\bf NO$_2$}, NO$_2^+$, {\bf HNO}, HNO$^+$, H$_2$NO$^+$, & 7 \\
(N-S) & NS, NS$^+$, HNS$^+$ & 3 \\
(O-H) & O, O$^+$, O$^{++}$, OH, OH$^+$, {\bf H$_2$O}, H$_2$O$^+$, H$_3$O$^+$, & 8 \\
(O-O) & {\bf O$_2$}, O$_2^+$, O$_2$H$^+$, & 3 \\
(O-S) & SO, SO$^+$, {\bf SO$_2$}, SO$_2^+$, HSO$_2^+$, & 5 \\
(S-H) & S, S$^+$, S$^{++}$, HS, HS$^+$, {\bf H$_2$S}, H$_2$S$^+$, H$_3$S$^+$, & 8 \\
(Si-H) & Si, Si$^+$, Si$^{++}$, SiH, SiH$^+$, SiH$_2$, SiH$_2^+$, SiH$_3$,
SiH$_3^+$, {\bf SiH$_4$}, SiH$_4^+$, SiH$_5^+$, & 12 \\
(Si-C) & {\bf SiC}, SiC$^+$, HCSi$^+$, & 3 \\
(Si-N) & {\bf SiN}, SiN$^+$, HNSi$^+$, & 3 \\
(Si-O) & {\bf SiO}, SiO$^+$, SiOH$^+$, & 3 \\
(Si-S) & {\bf SiS}, SiS$^+$, HSiS$^+$, & 3 \\
(Na) & Na, Na$^+$, Na$^{++}$, & 3 \\
(Mg) & Mg, Mg$^+$, Mg$^{++}$, & 3 \\
(Fe) & Fe, Fe$^+$, Fe$^{++}$, & 3 \\
(Ne) & Ne, Ne$^+$, Ne$^{++}$, & 3 \\
(Ar) & Ar, Ar$^+$, Ar$^{++}$, & 3 \\
\hline
species & total & 169 \\
\hline
\end{tabular}}
\tablefoot{All stable neutral molecules are marked in bold face.}
\end{table*}

Equation (\ref{kreverse}) implies that the reverse and forward rate coefficients are not independent and are related to each other via the reaction's Gibbs free energy.  Using this principle is standard in exoplanet chemistry, but is normally not used in interstellar cloud or disk chemistry. The concept can be applied not only to all reaction types listed in Sect.~\ref{3body}, but also to regular bimolecular neutral-neutral, ion-neutral, and ion-ion gas phase reactions that have no high-pressure limit data.  

Equation (\ref{kreverse}) can be reliably applied to the case where the forward reaction is an exothermic reaction, $\dGR\!<\!0$, and the reverse reaction rate coefficient will be small because of the negative term in the exponential in Eq.~(\ref{kreverse}). However, there are cases where the rate measurements are only available for an endothermic reaction (for example, a termolecular reaction) and the reverse exothermic reaction needs to be constructed. This is still safe as long as the measured activation energy, $\gamma_0$, in Eq.\,(\ref{Arrhenius}) is larger than the positive $\dGR$. However, if that is not the case, the constructed rate would go to infinity at low temperatures. This problem is further discussed in Sect.\,\ref{Implementation}.

\subsection{Thermochemical data}

The thermochemical data, such as the standard formation enthalpies, $\dH$, and the standard Gibbs free energies of formation, $\dG$, are calculated from NASA seven-term polynomials given by \citet{Burcat}. These consist of 14 coefficients. The first seven coefficients are used to determine the thermodynamic properties for temperatures 1000-6000\,K and the subsequent seven coefficients are used for 200-1000\,K. 

We carefully checked the standard heat of formation, $\dH(0\rm\,K)$, against the data provided by \citep{Millar1997}, the National Institute of Standards and Technology (NIST) Standard Reference Database Number 69\footnote{\url{https://webbook.nist.gov/chemistry}}, and the Active Thermochemical Tables (ATcT)\footnote{\url{https://atct.anl.gov/Thermochemical\ Data/version\ 1.130/index.php}}. In most cases, we find fine agreements between the four data sources, but also some disagreements.  About 30 molecules in our selection could not be found in \citet{Burcat} and \citet{Dick19}.  In these cases we (i) fitted the thermochemical data found in Gleich-Gewichts-Chemie (GGchem) and exported them in Burcat format, (ii) estimated the Burcat polynomials from similar molecules calibrated with the correct $\dH(0\rm\,K)$, as found in the ATcT and NIST databases, or (iii) used known ionisation potentials, or (iv) used proton affinities with respect to their mother molecules. The details are explained in Table~\ref{tab:newBURCAT}.

\section{Implementation}\label{Implementation}

Here we describe the implementation of the above-mentioned concepts in the chemical network of the thermochemical disk code P{\tiny RO}D{\tiny I}M{\tiny O} \citep{Woitke2009,Woitke2016}. 

\subsection{Selection of species} \label{Selection of species}

To study whether thermodynamic equilibrium can be established in disks we made a selection of species based on the large DIANA (DIscANAlysis) chemical network \citep{Kamp2017}. However, since we discuss rather high temperatures of $T\!\gtrsim\!500$\,K, we eliminated all ice species. We have furthermore eliminated excited molecular hydrogen, $\rm H_2^\star$, and polycyclic aromatic hydrocarbons (PAHs), due to the lack of thermochemical data for those, resulting in the set of 169 species listed in Table\,\ref{tab:species}.
All but the ten doubly ionised atoms have valid thermodynamic data. It is noteworthy that our selected species included atomic and molecular ions, which is not standard in exoplanet chemistry, to study the details of photoionisations and cosmic rays, and doubly ionised atoms to study the effects of X-ray processes implemented by \citet{Aresu2011}. We reintroduced the eliminated species later in Sect.\,\ref{Discussion_ch4} to study the 2D abundance structure in full disk models. 

\subsection{Reaction data and pairs of forward and reverse reactions}
\label{chemnetwork}

Our primary source of reaction kinetic data was the UMIST 2022 database \citep{Millar2024}. Additional gas-phase reactions were taken from the STAND network \citep{Rimmer2016}, providing the majority of three-body reactions with low- and high-pressure limits. The STAND reactions have preference in this work: that is, if the same reaction was found in both databases, we used the kinetic reaction data from STAND. This includes the radiative association and direct recombination reactions, but not the photoionization and photodissociation reactions, which are treated in a different way in {\sc ProDiMo}. Only reactions among the selected species were read from these databases. In STAND, the reactions are already paired. They come in the form of a reaction with valid Arrhenius coefficients followed by the reverse reaction with invalid data.  We ignored this pairing and simply added each reverse reaction with a flag signalling that the reaction rate of this reaction needed to be derived later from its respective forward rate, using Eq.\,(\ref{kreverse}). 

Next, each gas-phase reaction is checked whether a reverse reaction is present, in which case these two reactions are paired.  If an invertible reaction is found that has no reverse counterpart, the reverse reaction is auto-created with a flag that it has no valid data, then we pair these two reactions. However, a STAND reverse reaction, which has invalid reaction data, will not overwrite the same UMIST\,2022 reaction with valid data. A gas-phase reaction is considered to be invertible if their reactants and  products all have valid thermodynamic data.  However, gas-phase reactions with less than two reactants (spontaneous reactions) or more than three products are not inverted. Once the pairing of the reactions is completed, we check whether at least one reaction in each pair has valid kinetic data.  We also check whether any unpaired invalid reactions remain, in which case the program stops with an error message.

\begin{table}
    \caption{Additional neutral-neutral reactions for silicon hydrides added.}
    \vspace*{-2mm}
    \resizebox{\linewidth}{!}{%
    \begin{tabular}{cccccc} \hline
    Reactions & $\alpha$  & $\beta$ & $\gamma$ & T\,[K] limit \\
    &&&&\\[-2.1ex]
    \hline
    &&&&\\[-2.1ex]
    \ce{SiH} + \ce{H2} + M $\rightarrow$ \ce{SiH3} + M   & 1.59(-30) &  -2.96 & 813.00  &  300-2000 \\
    \ce{SiH2} + \ce{H2} + M $\rightarrow$ \ce{SiH4} + M   & 7.07(-29)  & -4.44  & 1202.00 & 300-2000 \\
    \ce{SiH} + H $\rightarrow$ \ce{Si} + \ce{H2}  & 6.59(-11) & -0.18 & 222.9 & 300-2000 \\
    \ce{SiH2} + H $\rightarrow$ \ce{SiH} + \ce{H2}  & 3.38(-11) & 1.82 & 100.0 & 300-2000  \\
    \ce{SiH3} + H $\rightarrow$ \ce{SiH2} + \ce{H2} & 2.67(-10) & -0.05 & 77.0 & 300-1000 \\ 
    \ce{Si} + \ce{N2O} $\rightarrow$ \ce{NO} + \ce{SiN} & 8.30(-10) & 0.0 & 8094.0 & 1780-3560\\
    \hline
    \end{tabular}} 
    \tablefoot{These reactions are taken from \cite{Raghunath2013} and NIST \citep{Manion2015}. $a(b)$ means $a\times10^b$.}
    \label{Si_table}
\end{table}

Altogether, the reaction network merged in this way consisted of 5539 (2462+375+1155+1547) reactions. Of these, 2462 reactions were from the UMIST 2022 database and 375 were taken from a collection of additional reactions compiled in the past by various {\sc ProDiMo} developers (see \citealt{Aresu2011}, \citealt{Kamp2017} and \citealt{Kanwar2023}), resulting in 2837 reactions. This compilation now also includes some additional three-body and neutral-neutral reactions of silicon hydrides, as listed in Table~\ref{Si_table}. The UMIST 2022 database only has ion-neutral reactions for silicon hydrides; see further discussion in Sect.\,\ref{Results}. There were 1155 reactions from the STAND network and 736 existing reactions were overwritten. Finally, 1547 reverse reactions were auto-created as discussed above, resulting in 2466 reaction pairs. A total of 75 gas-phase reactions out of 5539 reactions remained unpaired. 

\begin{table}
    \caption{Number of various types of reactions taken from the STAND network.} 
    \label{tab:types}
    \vspace*{-2mm}
    \label{types}
    \resizebox{\linewidth}{!}{%
    \begin{tabular}{ccc} \hline
    Type of reactions & \hspace*{-2mm} Number of reactions\hspace*{-5mm} & Comments\\ \hline
    &&\\[-2.1ex]
    A + B + M $\rightarrow$ AB + M     & 50 & termolecular\\
    AB + M $\rightarrow$ A + B + M     & 48 & thermal decomposition\\
    \ce{A+} + B + M $\rightarrow$ \ce{AB+} + M  & 5 & ion-termolecular \\
    \ce{AB+} + M $\rightarrow$ \ce{A+} + B + M  & 5 & ion-thermal decomposition  \\
    A + B $\rightarrow$ C + D  & 332 & neutral-neutral bimolecular\\
    C + D $\rightarrow$ A + B   & 235  & \hspace*{-2mm}reverse neutral-neutral bimolecular\hspace*{-2mm}\\
    \ce{A+} + B $\rightarrow$ \ce{C+} + D & 549 & ion-neutral bimolecular\\
    \ce{C+} + D $\rightarrow$ \ce{A+} + B & 463 & \hspace*{-2mm}reverse ion-neutral bimolecular\hspace*{-2mm}\\
    \ce{A+} + e$^{-}$ + M  $\rightarrow$ A + M  & 46  & 3-body recombination\\
    A + M $\rightarrow$ \ce{A+} + e$^{-}$ + M   & 46 & thermal ionization\\ 
    A$^+$ + \ce{e-} $\rightarrow$ \ce{A} + h$\nu$ & 5 & radiative association \\
    AB$^+$ + \ce{e-} $\rightarrow$ A + B  & 107 & dissociative recombination \\ \hline
    \end{tabular}}
    \tablefoot{A, B and C denote gas-phase species whereas M represents the third-body. The total number of these reactions are 1891 which is consistent with the sum of reactions added and overwritten by the STAND (1155+736).}
\end{table}

The final network included 135 X-ray driven reactions, 252 photoreactions, and 203 cosmic-ray driven reactions. The photoreactions here also included radiative association and direct recombination reactions. The cosmic-ray reactions included cosmic-ray particle reactions and ionisation by UV photons generated by the cosmic rays. Table\,\ref{tab:types} provides an overview of the different types of reactions picked up from the STAND network. The number of forward reactions included exceeds the number of reverse reactions because for some of the STAND reverse reactions with invalid reaction data the same reaction is found in UMIST 2022 with valid reaction data.

\subsection{Computation of the rate coefficients}
\label{sec:reverse_rates}

To calculate the rate coefficients of the paired reactions, we first determined the reaction enthalpy, $\dHR(T)$, from NASA seven-term polynomials to decide which one was the exothermic reaction.  If that reaction had valid data, it was considered as the "forward" reaction and the paired endothermic reaction was the reverse.  However, if the exothermic reaction had no valid rate coefficient data, it became the reverse reaction, and the endothermic reaction became the forward reaction. In both cases, the rate coefficient of the reverse rate was finally computed according to Eq.\,(\ref{kreverse}). If the exothermic reaction did not have reaction data and the endothermic reaction was identified as the forward one, we have $\dGR>0$ and Eq.\,(\ref{kreverse}) can potentially lead to $K_r\!\to\!\infty$ for $T\!\to\!0$.  Combining Eq.\,(\ref{kreverse}) with Eq.\,(\ref{Arrhenius}) for the forward rate, we have 
\begin{equation}
  K_{\rm r} = \biggl({\frac{\pst}{kT}}\biggl)^{N} 
  \alpha_{\rm f}\;\bigg(\frac{T}{300}\bigg)^{\beta_{\rm f}}
    \exp\left(+\frac{\dGR/R-\gamma_{\rm f}}{T}\right) \ .
    \label{correction}
\end{equation}
Thus, there is a problem when $\dGR/R>\gamma_{\rm f}$. This case seems chemically impossible, as an endothermic reaction should have an activation energy that is at least as large as the reaction enthalpy. However, in practice this case occurs frequently whenever an endothermic reaction has been reported with low or zero activation energy. \citet{Tinacci2023} identify this problem as one of the major weaknesses of current theoretical astrochemical modelling and propose a cleaning of chemical networks by simply eliminating all these spurious reactions.  Here, we followed the solution that \cite{Rimmer2016} propose, to multiply both $K_{\rm f}$ and $K_{\rm r}$ by the same factor $\exp\left(-\frac{\frac{\dGR}{R}\,-\,\gamma_{\rm f}}{T}\right)$ if $\dGR/R>\gamma_{\rm f}$, which not only avoids the problem above, but also reduces the rate coefficients of all spurious endothermic forward reactions automatically (auto-cleaning).

\subsection{At which densities do three-body reactions become important?}

Computing an average over all exothermic three-body reaction rates in the new merged chemical network at 1000\,K results in $\langle\log_{10}K^{\rm exotherm}_{\rm 3-body}\rm\,[cm^6s^{-1}]\rangle\!=\!-31.9\pm 4.8$, whereas the average over all exothermic two-body reaction rates is $\langle\log_{10}K^{\rm exotherm}_{\rm 2-body}\rm\,[cm^3s^{-1}]\rangle\!=\!-12.1\pm 3.4$. From this simple consideration one might be led to the conclusion that three-body reaction rates are only important when the density is as large as $10^{20}\rm\,cm^{-3}$. However, this conclusion is incorrect.  The termolecular reactions open new chemical pathways with reactions that are not possible without a stabilising collision, and in our chemical analysis we see 
three-body reactions being flagged as important once the density is of the order of $10^{13}\rm\,cm^{-3}$ or higher. The following consideration tries to explain this observation.

In order to assess the density beyond which the inclusion of the three-body (termolecular) reactions seems important, we searched in the network for pairs of three-body and two-body reactions that share the same reactants, except for the additional M in the three-body reaction. An example for such reactions is listed in Table~\ref{detailed_balance}: see the first two destruction reactions listed for \ce{C2H2}. From the rate coefficients of such a pair we can calculate a critical density for the third body as
\begin{equation}
    n_{\rm cr}^{\rm ter} = \frac{K_{\rm 2-body}}{K_{\rm 3-body}} \ .
\end{equation}
Each of these pairs of reactions means a branching point. Either the activated complex is stabilised by a collision with a third body, resulting in a constructive exothermic process, or it decays spontaneously into two fragments, often endothermically. In this way, the exothermic termolecular reactions often compete with the endothermic two-body reactions which have much lower rates than their exothermic counterparts. From Eq.\,(\ref{Lind}), one can derive another critical density beyond which the high-pressure limit becomes relevant:
\begin{equation}
    n_{\rm cr}^{\rm high-p} = \frac{K_\infty}{K_0} \ .
\end{equation}

Table~\ref{tab:meanrates} lists the mean values we find for these two critical densities in our network, at various temperatures. The critical density for the inclusion of termolecular reactions, $n_{\rm cr}^{\rm ter}$, shows a broad distribution among the molecules, but values from roughly $10^{11}\rm\,cm^{-3}$ to $10^{15}\rm\,cm^{-3}$ are often found. 
The values are lower at low temperatures, where the termolecular reactions are often faster than their endothermic two-body counterparts that have higher activation barriers.  In contrast, the critical density for the inclusion of high-pressure data varies from about $10^{17}\rm\,cm^{-3}$ to $10^{22}\rm\,cm^{-3}$, depending on reaction and temperature, which seems unlikely to be important for any region in protoplanetary disks. Thus, including the three-body termolecular reactions seems important for the inner disks, but using only the low-pressure reaction data is justified. \cite{Kamp2017} and \cite{Kanwar2023} already include a limited set of three-body reactions; however, the network developed here is more exhaustive for three-body reactions. 

\begin{table}
   \caption{Critical densities beyond which termolecular reactions and their high-pressure limits become important.}
   \label{tab:meanrates}
   \vspace*{-2mm}
   \begin{tabular}{c|cc}
   \hline
   &&\\[-2.1ex]
          & $\langle \log_{10}\,n_{\rm cr}^{\rm ter}\rangle\rm\,[cm^{-3}]$ 
          & $\langle \log_{10}\,n_{\rm cr}^{\rm high-p}\rangle\rm\,[cm^{-3}]$\\
   &&\\[-2.1ex]
   \hline
   &&\\[-2.1ex]
   200\,K & $12.7\pm 7.0$ & $19.2\pm 2.1$ \\
   400\,K & $13.5\pm 6.0$ & $19.2\pm 2.1$ \\
   1000\,K& $12.2\pm 6.9$ & $19.2\pm 1.0$ \\
   2000\,K& $16.2\pm 4.1$ & $19.1\pm 1.2$ \\ 
   \hline
   \end{tabular}
\end{table}

\begin{table*}
\centering
\caption{Model grids explored in Sect.~4 to test the {\sc ChaiTea} chemical network.}
\label{density}
\vspace*{-2mm}
\begin{tabular}{|l|cccc|}
\hline 
&&&&\\[-2.2ex]
Hydrogen nuclei density $\nH$ [cm$^{-3}$] & 
\multicolumn{4}{|l|}{5.546$\times$10$^{10}$, 7.688$\times$10$^{12}$, 7.562$\times$10$^{13}$, 6.855$\times$10$^{14}$, 4.770$\times$10$^{15}$, 8.296$\times$10$^{16}$} \\ 
&&&&\\[-2.2ex]
\hline
&&&&\\[-2.2ex]
Temperature [K]                           & \multicolumn{4}{l|}{\begin{tabular}[c]{@{}l@{}}1500, 1450, 1400, 1350, 1300, 1250, 1200, 1150, 1100, 1050, 1000, 950,\\ 900, 850, 800, 750, 700, 650, 600, 550, 540, 530, 525, 520, 515, 510, 500\end{tabular}}\\ 
\multicolumn{5}{|c|}{}\\[-2.2ex]
\hline
\multicolumn{5}{|c|}{}\\[-2.2ex]
\multicolumn{5}{|c|}{Grids}\\ 
\multicolumn{5}{|c|}{}\\[-2.2ex]
\hline
Grid names & Photo-processes & X-rays   & CR ionisation rate [s$^{-1}$] & \# of models \\ 
&&&&\\[-2.2ex]
\hline
&&&&\\[-2.2ex]
grid\_0     & $\times$     & $\times$ & $\times$ & 162\\
grid\_photo & $\checkmark$ & $\times$ & $\times$ & 162\\
grid\_cr    & $\times$ & $\times$ & 
\begin{tabular}[c]{@{}l@{}}1.7$\times$10$^{-40}$, 1.7$\times$10$^{-35}$, 1.7$\times$10$^{-30}$, \\
1.7$\times$10$^{-25}$, 1.7$\times$10$^{-20}$, 1.7$\times$10$^{-17}$\end{tabular} & 972 \\ 
\hline       
\end{tabular}
\end{table*}

\subsection{Disk modelling}\label{Disk modelling}

We used the thermochemical disk code P{\tiny RO}D{\tiny I}M{\tiny O} \citep{Woitke2009,Woitke2016} to establish in which parts of the disk thermochemical equilibrium can be established. This code first sets up an axisymmetric physical gas and dust density structure, and then calculates the 2D continuum radiative transfer to determine the dust temperature structure, $T_{\rm dust}(r,z)$. 
{\sc ProDiMo} then iteratively calculates the chemical and thermal structure of the gas in the disk using the enhanced chemical network described in Sect.\,\ref{chemnetwork}. The gas temperature is varied at every point in these calculations until the local gas heating rate is balanced by the local gas cooling rate, taking into account heating and cooling processes \citep[see][]{Woitke2015}. The treatment of cosmic rays follows the approach described by \cite{Padovani2009}, \cite{Padovani2013} and \citet[see Eq.\,(4)]{Rab2017}.

As a basis for all modelling steps to follow, we computed a standard T\,Tauri {\sc ProDiMo} disk model adopted from \citet{Woitke2016} with the enhanced chemical network described above. The full list of input parameters of the disk model are provided in Appendix\,\ref{disk_parameters}. The elemental abundances were adopted from \cite{Kamp2017}. The C/O elemental ratio used in this work is 0.45. 

\section{Results}\label{Results}

\subsection{Verification of the chemical network}\label{Verification of the chemical network}

Finally, we obtain a chemical network that we named CHemistry Assembled from exoplanets and dIsks for Thermodynamic EquilibriA ({\sc ChaiTea}), which consists of the UMIST 2022, large DIANA, and STAND network. Our first goal was to verify whether the {\sc ChaiTea} chemical network was able to converge towards thermodynamic equilibrium. We checked this by selecting a single point from the T\,Tauri disk model, with given hydrogen nuclei density, $\nH$, and only simulating the chemistry at that point either in the time-dependent or in the time-independent mode.  For the purpose of this test, we eliminated all possible causes for deviations from thermodynamic equilibrium:
(i)\,we switched off the cosmic-ray and X-ray reactions by using extremely low values for the cosmic-ray ionisation rate and the stellar X-ray luminosity, (ii)\,we set the dust and gas temperatures equal to a user-specified value $T_{\rm gas}\!=\!T_{\rm dust}\!=\!T$ and, (iii)\,we set the local radiation field to a Planckian $J_\nu\!=\!B_\nu(T)$, (iv)\, we switched off all photo rates including any reactions that produce photons, such as radiative associations and direct recombinations. We defined two conditions to claim that the chemistry had reached thermodynamic equilibrium: (1) the forward and reverse rates of all reaction pairs are equal (detailed balance) and (2) the calculated abundances are equal to the abundances computed with an independent code that predicts them in thermodynamic equilibrium, $n_i\!=\!\neqq_i$. For the latter, we used GGchem \citep{Woitke2018} with a new option to use the same Burcat thermochemical data as used in this work. For our selection of 12 elements, GGchem finds 1450 molecules in its databases, including many isomers and molecules as complex as \ce{[(O2N)3C6H2CH]2} (hexanitrostilbene). However, for all temperatures considered in this paper, these complex molecules only attain trace abundances.  

We produced three grids of models with varying density and temperature. These models and their parameters are outlined in Table\,\ref{density}. All three grids explore the same density and temperature range. First, we produced grid\_0, where X-rays, the photoreactions and cosmic rays are turned off. Second, we studed the deviations caused by photoreactions based on a local Planck field $J_\nu=B_\nu(T)$ and reactions producing photons, such as radiative association and direct recombination reactions that are considered non-invertible, leading to another grid of 6$\times$27 models called grid\_photo. Third, to discern the impact of the cosmic rays, we again switched off the photorates and produced a third grid of models for six different cosmic-ray ionisation rates (CRIs), leading to grid\_cr with 6$\times$6$\times$27 models.

\begin{table}[]
\centering
\caption{Formation and destruction of \ce{C2H2} in kinetic chemical equilibrium at $T\!=\!1200$\,K and $\nH\!=\rm\!7.56\times10^{13}\,cm^{-3}$.}
\label{detailed_balance}
\vspace*{-2mm}
\resizebox{80mm}{!}{
\begin{tabular}{ll}
\hline
&\\*[-2.2ex]
formation reactions     & rates\,$\rm[cm^{-3}s^{-1}]$\\ 
\hline
&\\*[-2.2ex]
\ce{C2H} + \ce{H2} $\rightarrow$ \ce{C2H2} + H        & 1.60(-13) \\
\ce{C2H3} + \ce{M} $\rightarrow$ \ce{C2H2} + H + M    & 1.01(-14) \\
\ce{C2H3} + H $\rightarrow$ \ce{C2H2} + \ce{H2}       & 2.20(-16) \\
\ce{C2H} + \ce{H2O} $\rightarrow$ \ce{C2H2} + \ce{OH} & 4.12(-17) \\
\ce{CH3} + CO $\rightarrow$ \ce{C2H2} + OH            & 2.94(-18) \\
\ce{C2H4} + M $\rightarrow$ \ce{C2H2} + \ce{H2} + M   & 1.35(-18) \\
&\\*[-2.2ex]
\hline
&\\*[-2.2ex]
destruction reactions   & rates\,$\rm[cm^{-3}s^{-1}]$\\ 
\hline
&\\*[-2.2ex]
\ce{C2H2} + H  $\rightarrow$ \ce{C2H} + \ce{H2}       & 1.60(-13) \\
\ce{C2H2} + H + M $\rightarrow$ \ce{C2H3} + \ce{M}    & 1.01(-14) \\
\ce{C2H2} + \ce{H2} $\rightarrow$ \ce{C2H3} + H       & 2.20(-16) \\
\ce{C2H2} + \ce{OH} $\rightarrow$ \ce{C2H} + \ce{H2O} & 4.12(-17) \\
\ce{C2H2} + OH $\rightarrow$ CO + \ce{CH3}            & 2.94(-18) \\
\ce{C2H2} + \ce{H2} + M $\rightarrow$ \ce{C2H4} + M   & 1.35(-18) \\ 
&\\*[-2.2ex]
\hline
&\\*[-2.1ex]
particle density in chemical kinetics\,$n_{\ce{C2H2}}$ 
   & $7.272\times10^{-8}\rm\,cm^{-3}$\\[-1.1mm]
in thermodynamic~equilibrium~~$\neqq_{\ce{C2H2}}$\hspace*{-3mm}
   & $7.273\times10^{-8}\rm\,cm^{-3}$\\
total destruction/formation rate 
   & $1.7\times10^{-13}\rm\,cm^{-3}s^{-1}$ \\
\ce{C2H2} relaxation timescale & 0.014\,yr \\ \hline
\end{tabular}}
\tablefoot{All reactions are sorted according to their rates, showing the validity of detailed balance.}
\end{table}

\begin{figure}
   \centering
   \includegraphics[width=\linewidth]{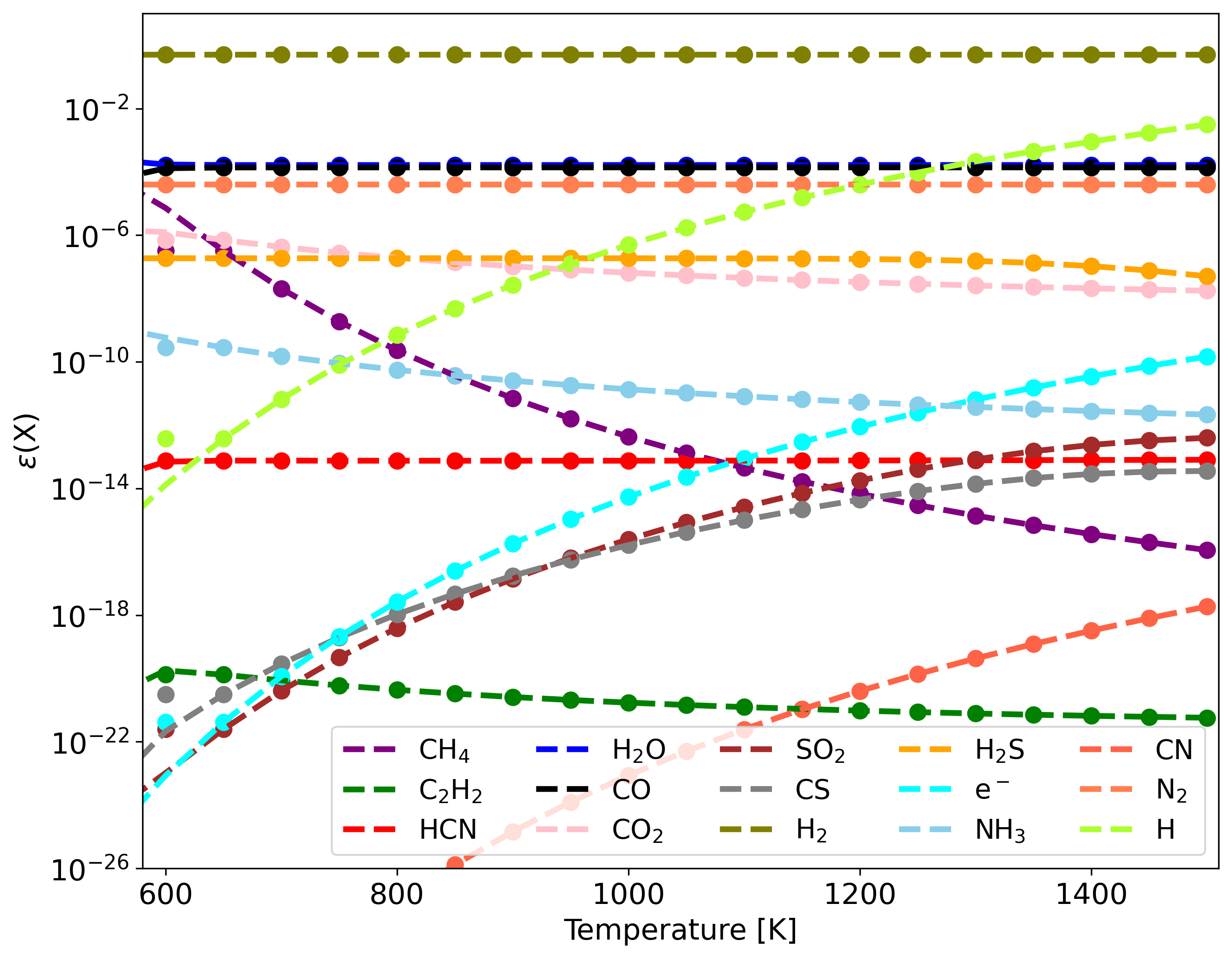}
   \caption{Abundances of various species, $\epsilon(X)\!=\!n_X/\nH$, obtained with the time-independent {\sc ProDiMo} models from grid\_0 for density $\nH\!=\!7.56\times 10^{13}\rm\,cm^{-3}$ (dots), compared to the corresponding GGchem models (dashed lines). The deviations at the lowest temperatures are due to numerical problems in converging the {\sc ProDiMo} models.}
  \label{TE13}
\end{figure}

\begin{figure*}[h!]
  \centering
  \hspace*{5mm} $T=1200$\,K \hspace*{75mm} $T=2000\,$K\\
  \includegraphics[width=\linewidth,trim=15 10 15 10,clip]{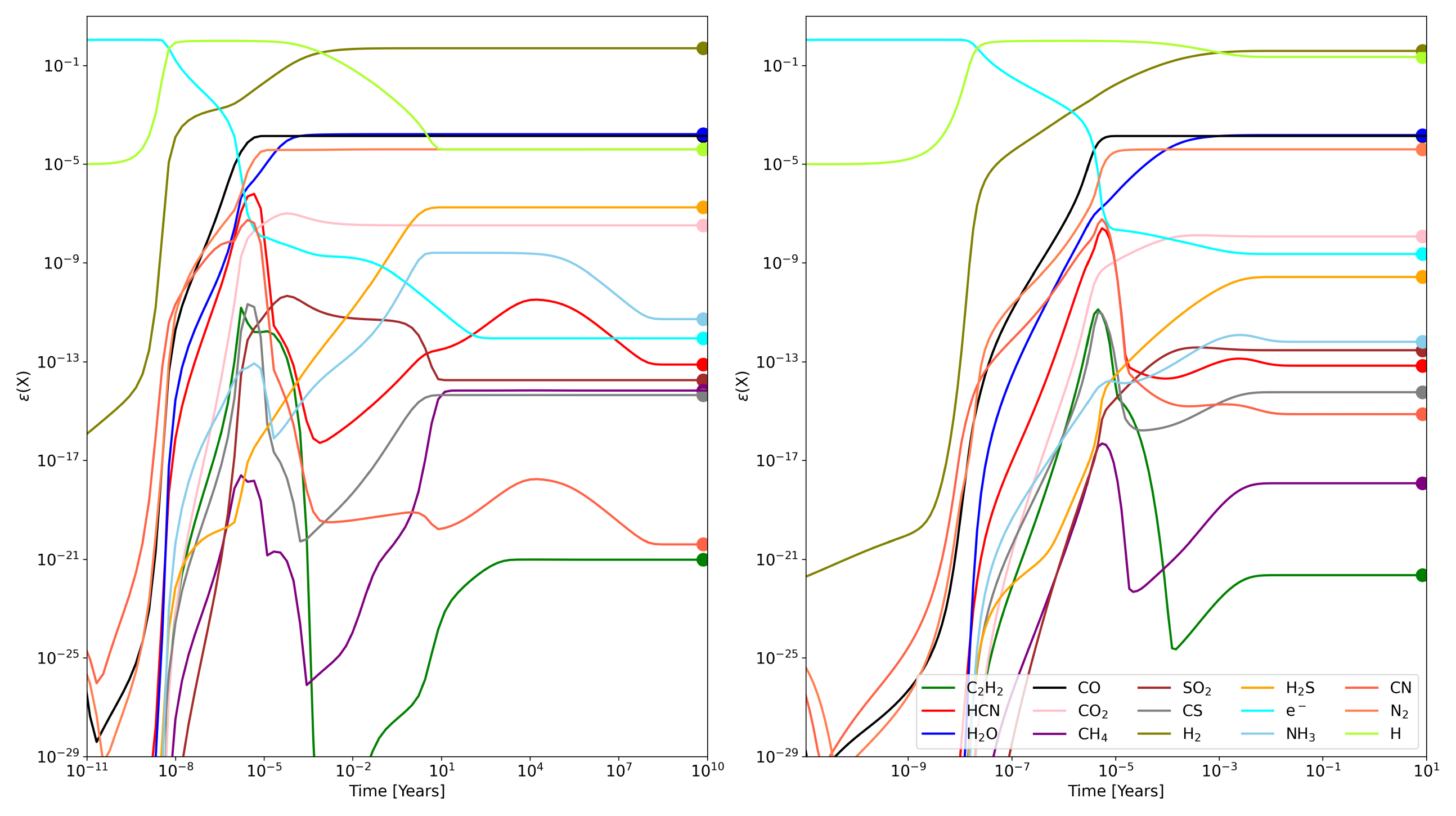}
  \caption{Abundances, $\epsilon(X)\!=\!n_X/\nH$, of various species (X) as a function of time obtained with P{\tiny RO}D{\tiny I}M{\tiny O} at 1200\,K and 2000\,K at a density of $\nH\!=\!7.56\times 10^{13}\rm\,cm^{-3}$. The large coloured dots represent the abundances in thermodynamic equilibrium obtained with GGchem.}
  \label{time-depend}
\end{figure*}

The {\sc ChaiTea} chemical network converges towards thermodynamic equilibrium in the absence of radiation such as cosmic rays, X-ray, and UV. We arrive at this conclusion irrespective of the molecule considered. Table\,\ref{detailed_balance} shows the detailed balance obtained for \ce{C2H2} solving the rate network in the time-independent mode at $T\!=\!1200$\,K and $\nH\!=\!7.56\times 10^{13}\rm\,cm^{-3}$ as part of grid\_0. It also shows that species such as \ce{H2O} and OH play a role in the formation of \ce{C2H2}, similarly to the findings of \cite{Kanwar2023}. Figure\,\ref{TE13} shows the calculated abundances of various species, determined with both {\sc ProDiMo} and GGchem, at a density of $\nH\!=\!7.56\times 10^{13}\rm\,cm^{-3}$.  These results match well for temperatures > 550\,K; below that, they begin to deviate from each other. These deviations are due to numerical problems in finding the correct chemical solution with the time-independent solver used in {\sc ProDiMo}. However, in general, our results show that the {\sc ChaiTea} chemical network will indeed approach thermodynamic equilibrium when all causes to deviate from it are eliminated.

\subsection{Chemical relaxation timescale}

We now analyse the chemical relaxation timescale ($\tau_{\rm chem}$) in the disk where the kinetic chemistry approaches thermodynamic equilibrium. This can tell us whether thermodynamic equilibrium can be established on disk-evolutionary timescales.
We realise that there can be various definitions for chemical relaxation timescales and explore them here. We define the chemical relaxation timescale in the following two different ways:
\begin{eqnarray}
    \tau_{\rm chem,1} &\!\!=\!\!& \max_{{\rm valid}\,i}\Big\{n_i/\min(F_i,D_i)\Big\} \label{1stm}\\
    \tau_{\rm chem,2} &\!\!=\!\!& \max_{{\rm valid}\,n}\big|\Re\{\lambda_n\}^{-1}\big| \label{2ndm} \ ,
\end{eqnarray}
where $F_i$ and $D_i$ are the total formation and destruction rates, respectively, of species $i$. $\tau_{\rm chem,1}$ can be computed from the diagonal elements in the chemical Jacobian. The slowest species determines the relaxation timescale. Equation\,(\ref{1stm}) is easy to understand but lacks a proper mathematical foundation. Using perturbation theory and linear algebra, the proper mathematical formulation of the relaxation timescale involves a transformation of the Jacobian into $n$ eigenvectors (the chemical modes) which each relax on individual timescales given by the inverse of the real part of their eigenvalues, $\lambda_n$. The slowest of these modes determines the chemical relaxation timescale, $\tau_{\rm chem,2}$, according to Eq.\,(\ref{2ndm}), \citep[see][]{Woitke2009}, with values listed in Table\,\ref{tab:relax}. Since a chemical network obeys a number of auxiliary conditions in the form of the conservation of the elements, the major element carriers must not be considered in Eq.\,(\ref{1stm}), and the $N_{\rm el}$ slowest modes must be discarded in Eq.\,(\ref{2ndm}), where $N_{\rm el}$ is the number of elements.

Figure\,\ref{time-depend} depicts the evolving abundances of various species in two time-dependent chemical models at 1200\,K and 2000\,K using a density of 7.65$\times$10$^{13}$\,cm$^{-3}$. We can visually read from this figure that the {\sc ProDiMo} abundances only approach the GGchem values after $\sim\!10^8$\,yr ($\tau_{\rm{chem,3}}$) in the 1200\,K model, but already approach them after $\sim\!0.1$\,yr in the 2000\,K model. These examples show how strongly the chemical relaxation timescale increases with decreasing temperature. 
\begin{table}
    \centering
    \caption{Chemical relaxation timescales obtained with different methods for 1200\,K and 2000\,K models at $\nH\!=\!7.5\times 10^{13}\rm \,cm^{-3}$, using time-dependent (t-dep.) and time-independent (t-indep.) chemistry.}
    \label{tab:relax}
    \vspace*{-2mm}
    \resizebox{\linewidth}{!}{
    \begin{tabular}{cc|c|c} 
    \hline
    $\tau_{\rm chem}$-method & \!\!chemistry\!\! & 1200\,K model & 2000\,K model \\ 
    \hline
    &&&\\[-2.2ex]
    $\tau_{\rm chem,1}$; Eq.\,(\ref{1stm})\!\!\! &  t-indep.
                                      & $1.68\times10^4$\,yr 
                                      & $1.28\times10^{-2}$\,yr \\
    $\tau_{\rm chem,2}$; Eq.\,(\ref{2ndm})\!\!\! & t-indep. 
                                      & $4.03\times10^7$\,yr  
                                      & 8.35\,yr \\[-1mm]
    $\tau_{\rm chem,3}$; visually $n_i \rightarrow \neqq_i$ & t-dep. 
                                      & $\sim\!10^8$\,yr 
                                      & $\sim\!0.1$\,yr\\ 
    \hline
    \end{tabular}}
\end{table}
While all three timescales at 2000\,K are relatively similar, the timescale at 1200\,K from the first definition ($\tau_{\rm chem,1}$) is about four orders of magnitude lower than $\tau_{\rm chem,2}$ and the visually determined value from the figure, which are both similar.

Hence, $\tau_{\rm chem,1}$ can be misleading. For example, proton exchange reactions between \ce{N2} and \ce{N2H+} have high rates, yet these reactions do not contribute to the formation and destruction of the strong N$\equiv$N bonds. Therefore, studying the chemical relaxation timescale of the chemical modes (eigenvectors of the Jacobian) is more reliable than the first definition. However, even that method does not predict the chemical relaxation timescale fully satisfactorily in all cases, especially when we consider even lower temperatures where the chemical relaxation timescale can easily become greater than the age of the universe. One of the possible explanations is that there are numerical problems when decomposing the chemical Jacobian into its eigenvectors. This is due to the extremely high conditional numbers of the matrix. Thus, from this analysis, we conclude that it is not trivial to reliably evaluate the chemical relaxation timescale that can have a huge range of fast and slow modes.

During this analysis, we find that species such as \ce{SiH} take extremely long to reach their equilibrium abundances based on a preliminary network. This indicates that this preliminary chemical network is incomplete for silicon-hydrides, as both UMIST~2022 and STAND only include reactions of Si-hydrides with charged species but no neutral-neutral reactions. We therefore added a few reactions from \cite{Raghunath2013} and NIST \citep{Manion2015}, as listed in Table\,\ref{Si_table}, to the {\sc ChaiTea} chemical network. This leads to a much faster relaxation of the Si-hydride molecules. However, once this problem is solved, the next molecules that present this issue are N-bearing species, which is further discussed in the following section. This might indicate that the network for these species is also not yet complete.

\subsection{The \ce{N2} formation problem}\label{N2_formation}

Figure\,\ref{time-depend} shows that N-bearing species such as \ce{HCN}, \ce{NH3}, and \ce{CN} are the slowest to attain thermodynamic equilibrium at 1200\,K, which happens only after $\sim$10$^8$ years. We analyse here the time-dependent behaviour at various steps for the two temperatures 1200\,K and 2000\,K. We started our simulations by putting all elements into their highest ionisation state as an initial condition, except for neutral hydrogen, which has an initial abundance of $10^{-5}$.  The gas recombines within  10$^{-8}$\,yr (i.e.\ in less than one second) at both temperatures. After about 10$^{-6}$\,yr ($\sim$30\,s), neutral molecules form by consuming the neutral atoms via barrier-free, exothermic neutral-neutral and termolecular reactions such as
\begin{eqnarray}
   \ce{H} + \ce{H} + \ce{H} \rightarrow \ce{H2} + \ce{H}\\
   \ce{H2} + \ce{O} \rightarrow \ce{OH} + \ce{H}\\
   \ce{OH} + \ce{N} \rightarrow \ce{NO} + \ce{H}\\
   \ce{NO} + \ce{N} \rightarrow \ce{N2} + \ce{O}.
\end{eqnarray}
This phase results in the creation of many simple neutral molecules that all attain similar abundances, as they are all produced by almost equally fast reactions.  Most of the atoms are consumed after about 0.1\,yr (one month) at 1200\,K and about 0.0001\,yr (one hour) at 2000\,K. At this time, the chemistry slows down substantially. Some parts of the network, such as the hydrocarbon chemistry, have already reached internal detailed balance, but as these parts are only weakly connected to other parts of the network, such as the N- and O-chemistry, the connections have not yet reached equilibrium. Eventually, after about 0.1\,yr in the 2000\,K model, each molecule has reached thermodynamic equilibrium. However, in the 1200\,K model, after 10$^3$\,yr all molecules have reached thermodynamic equilibrium except for the N-bearing species, which take as long as $10^8$\,yr to reach equilibrium. 
This final phase is characterised by difficulties in forming the thermodynamically favoured \ce{N2} from the other overabundant N molecules. All the reactions stated above become extremely inefficient due to the lack of free atoms at 1200\,K. The following reaction is found to be most effective in reducing the N-bearing species and forming N$\equiv$N bonds:
\begin{equation}
  \ce{NO} + \ce{NH2} \rightarrow \ce{N2} + \ce{H2O} \ ,
\end{equation}
but since both NO and \ce{NH2} are not very abundant molecules in thermodynamic equilibrium, this reaction channel takes as long as $10^8$\,yr in this example.

This analysis indicates that we may miss some other neutral-neutral reaction pathways to form or destroy N$\equiv$N bonds, in order to reduce the chemical relaxation timescale. We experimented with adding \ce{N2H}, \ce{NCN}, and \ce{HNCO} to our network, which indeed leads to a slight decrease in the relaxation timescales, but only by a factor of 2-3, and we are unable to solve the general problem this way. One possibility might be to include surface reactions to form \ce{N2} on much shorter timescales. At temperatures $\la\!1100\,$K, the chemical relaxation timescale derived from Eq.\,(\ref{2ndm}) becomes longer than the age of the universe. This issue was also reported by \cite{Gal2014}. The neutral-neutral reactions forming \ce{N2} mentioned by \cite{Pineau1990} and \cite{Hily2010} are included in the chemical network but are not found to be dominant.

\subsection{Deviations from thermodynamic equilibrium}

\begin{figure}
   \centering
   \includegraphics[width=\linewidth]{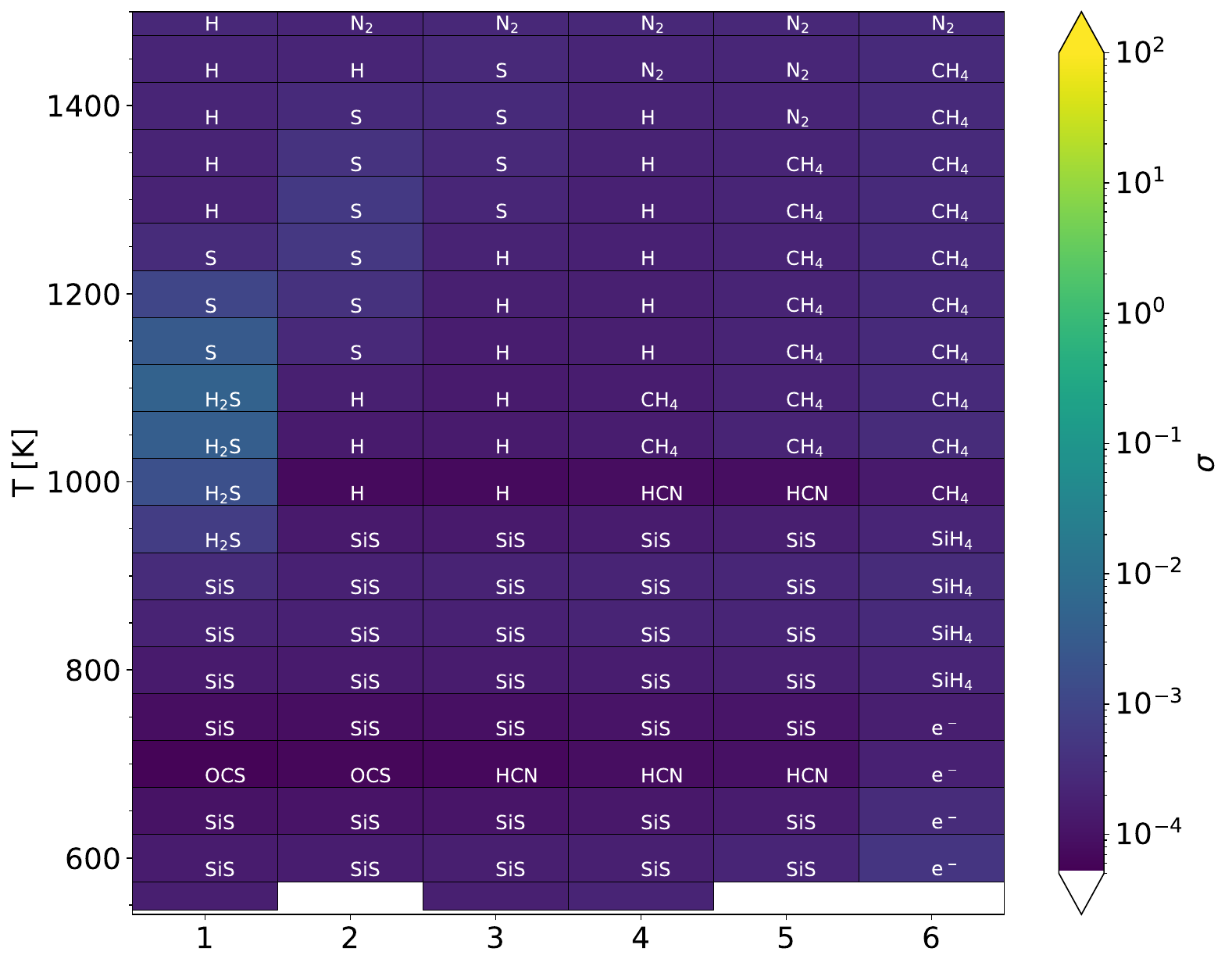}\\[-1mm]
   \# (density)
   \caption{Deviation $\sigma$ (see Eq.\,\ref{eq:sigma}), between particle densities obtained with time-independent {\sc ProDiMo} ($n_i$) and GGchem ($\neqq_i$) at various temperatures and densities from grid\_0. The blank rectangles at the bottom mark the models which do not converge. The numbers on the $x$-axis correspond to the ascending list of densities in Table\,\ref{density}. The central white labels mark the species that contribute most to $\sigma$.}
  \label{without_photo}
\end{figure}

Figure\,\ref{without_photo} depicts the deviations from thermodynamic equilibrium at different temperatures and densities when photoprocesses, X-rays, and cosmic rays are switched off (grid\_0). We define the deviation as
\begin{equation}
    \sigma = \sqrt{\frac{\sum_{i=1}^{N}
    \Big(\log_{10}n_i-\log_{10}\neqq_i\Big)^2\,w_i}
    {\sum_{i=1}^{N} w_i}},
    \label{eq:sigma}
\end{equation}
where $n_i$ is the {\sc ProDiMo} particle density and $\neqq_i$ is the GGchem particle density. $N$ is the number of species found in both models. To avoid picking up differences between particle densities at extremely low abundances such as $10^{-80}$, which {\sc ProDiMo} does not predict properly, we used a summation weight of $w_i\!=\!\big(\max\{n_i,\neqq_i\}/\nH\big)^{0.2}$. In this way, a deviation of one order of magnitude at an abundance of $10^{-20}$ is evaluated as as poor as a deviation of $10^{-4}$ at unit abundance. The mean deviation, $\sigma$, varies across temperature and density. We excluded \ce{Mg}, \ce{Mg+}, \ce{Fe}, \ce{Fe+}, \ce{Na}, and \ce{Na+} in Eq.\,(\ref{eq:sigma}) because the selection of Mg-, Na- and Fe- bearing species is incomplete in {\sc ProDiMo}, causing large deviations at low temperatures where GGchem predicts these metals to be present in the form of hydroxides such as \ce{Mg(OH)2}, \ce{Fe(OH)2}, and \ce{(NaOH)2}, respectively. 

\begin{figure}
   \centering
   \includegraphics[width=\linewidth]{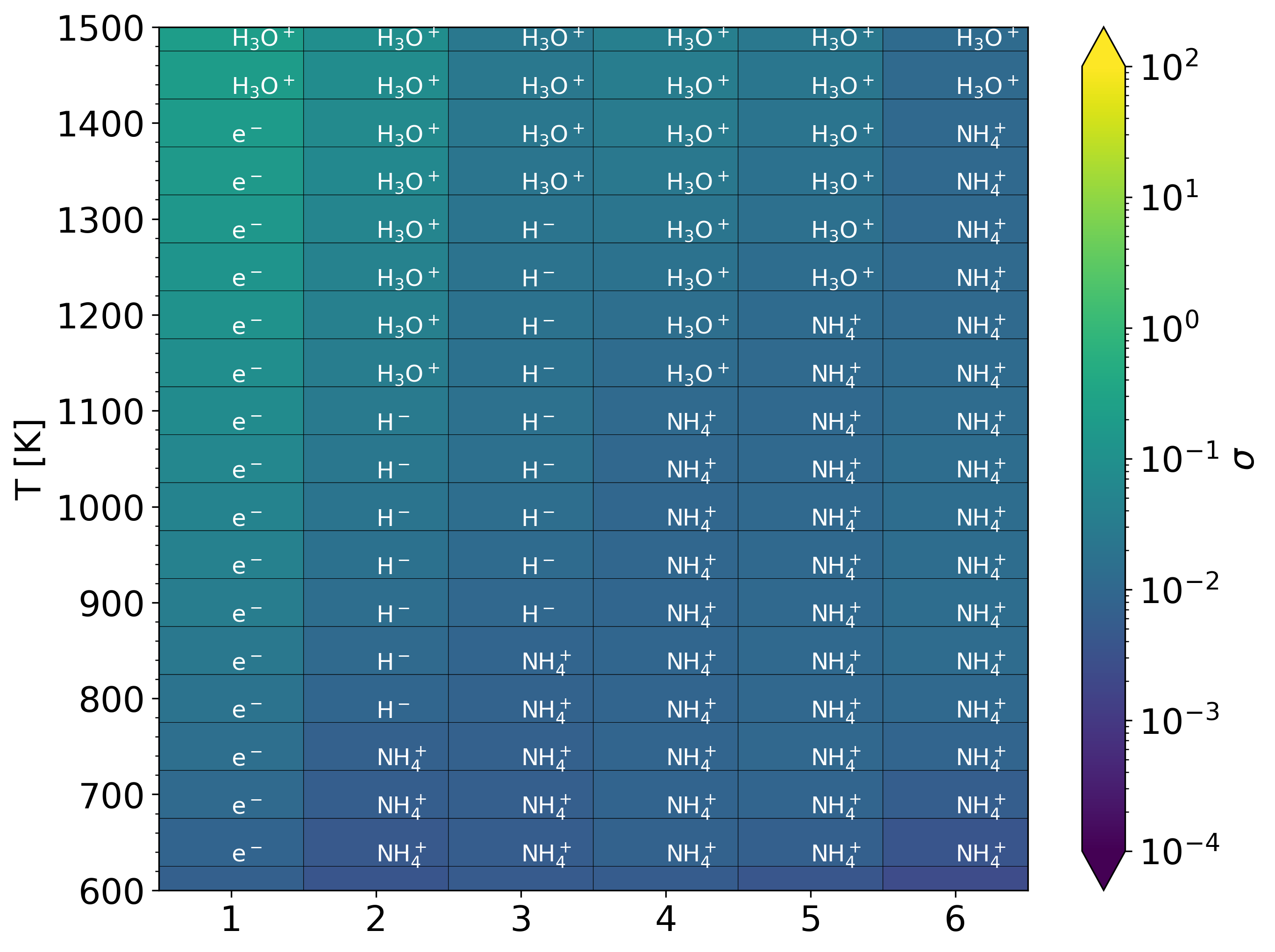}\\[-1mm]
   \# (density)
   \caption{Same as Fig.~\ref{without_photo} but from grid\_photo which includes photo processes based on a local Planck field $J_\nu=B_\nu(T)$, radiative association, and direct recombination reactions.}
  \label{with_photo}
\end{figure}

\begin{figure*}
\begin{tabular}{cc}
\includegraphics[width=0.5\textwidth]{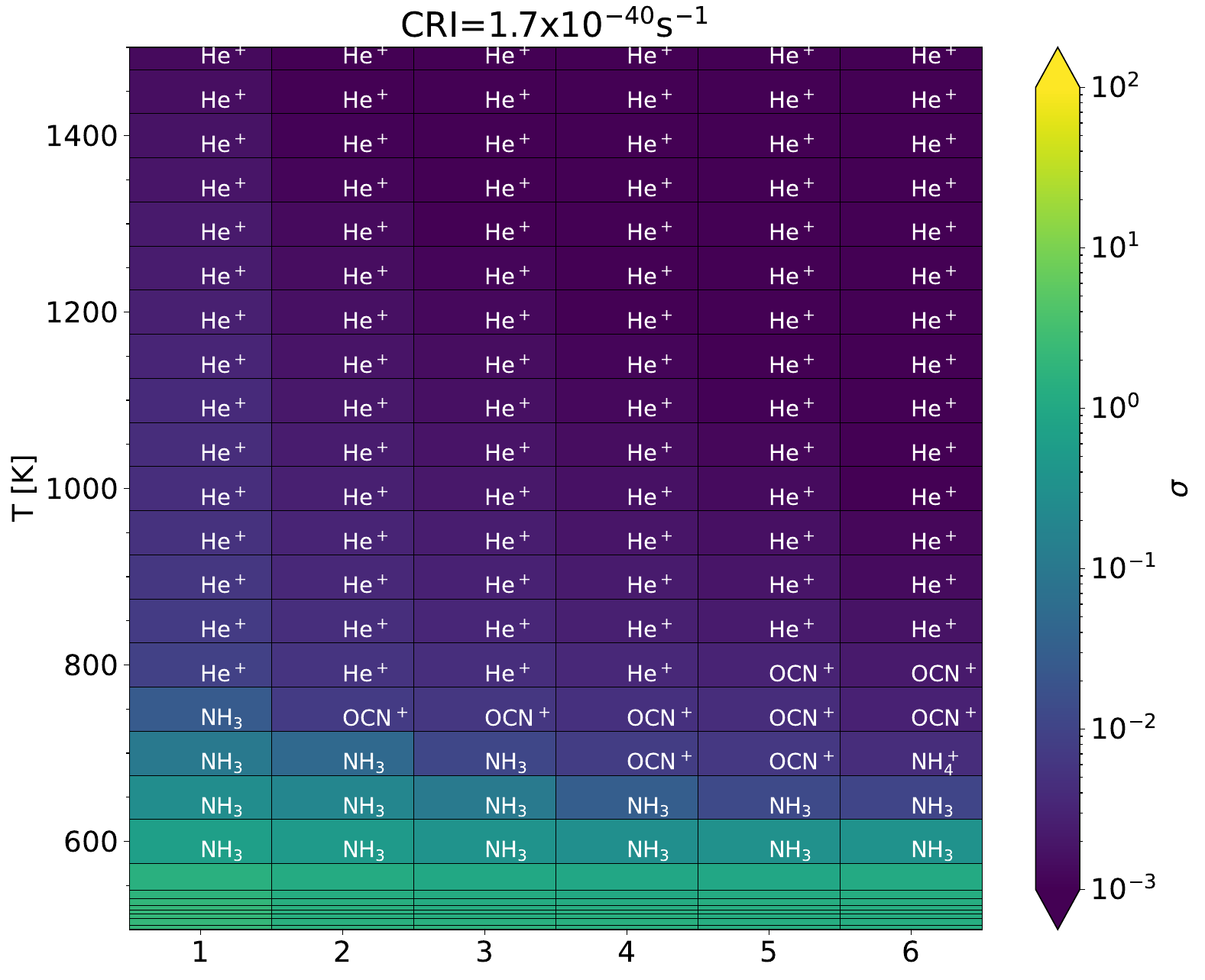}
&
\includegraphics[width=0.5\textwidth]{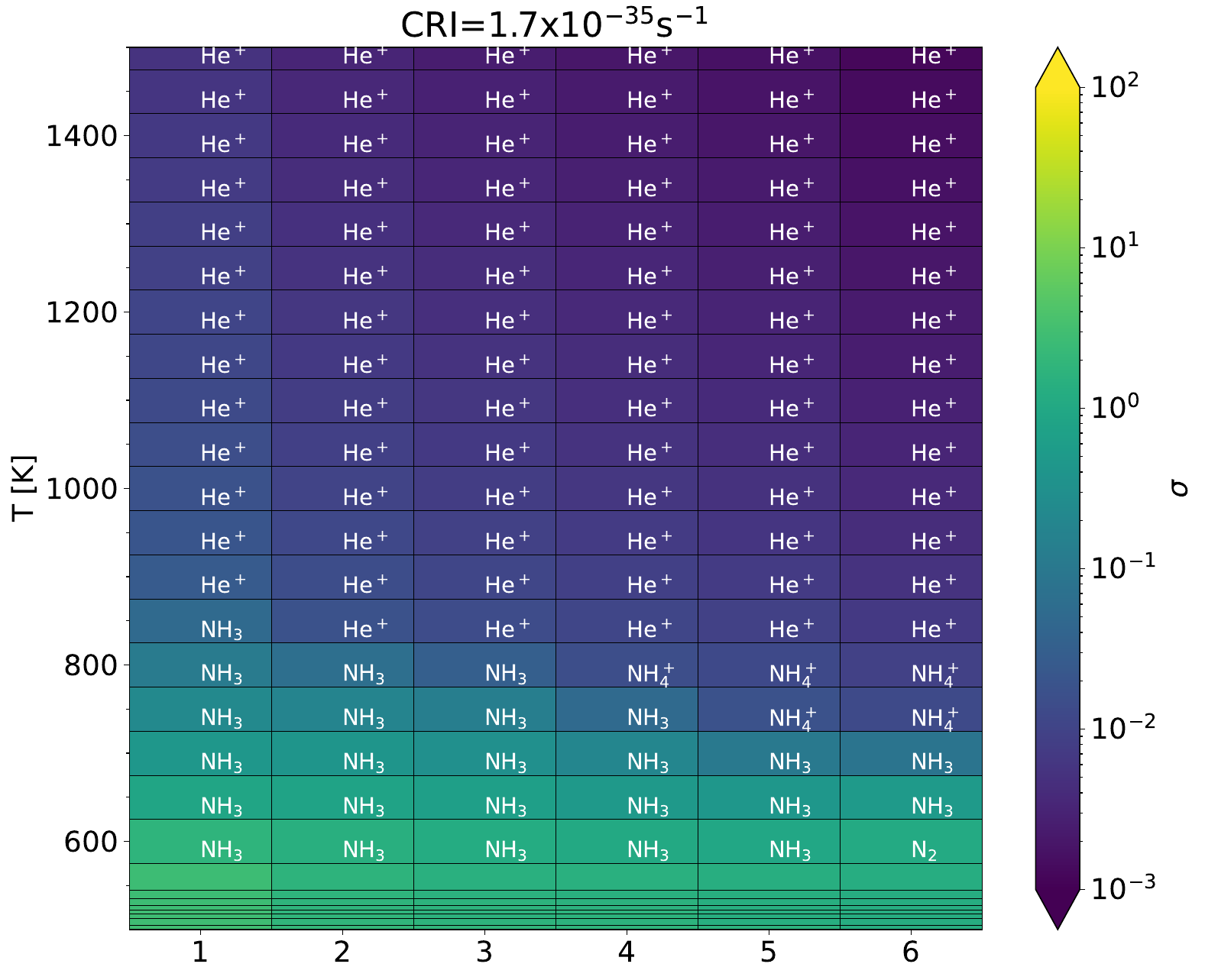}
\\
\includegraphics[width=0.5\textwidth]{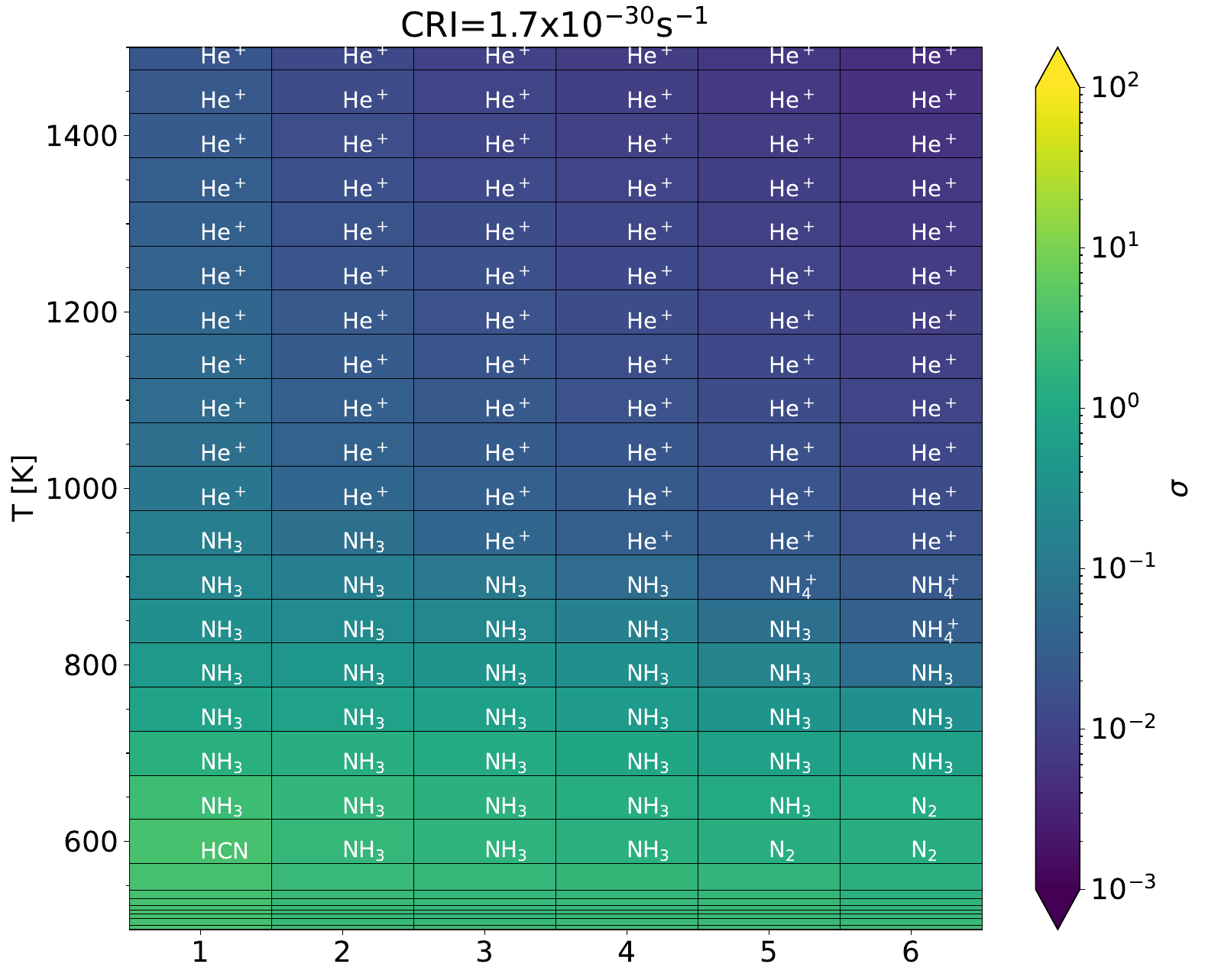}
&
\includegraphics[width=0.5\textwidth]{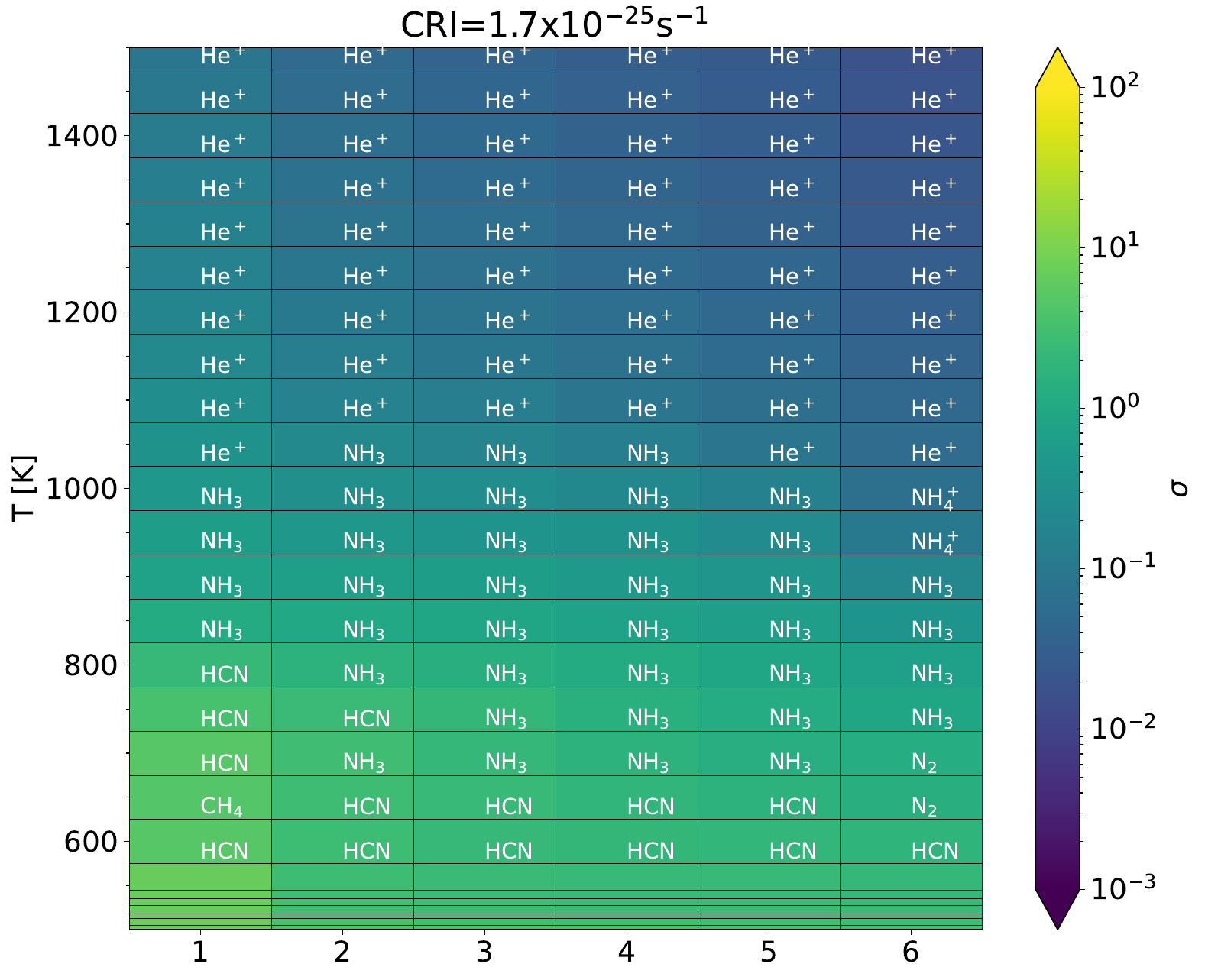}
\\
\includegraphics[width=0.5\textwidth]{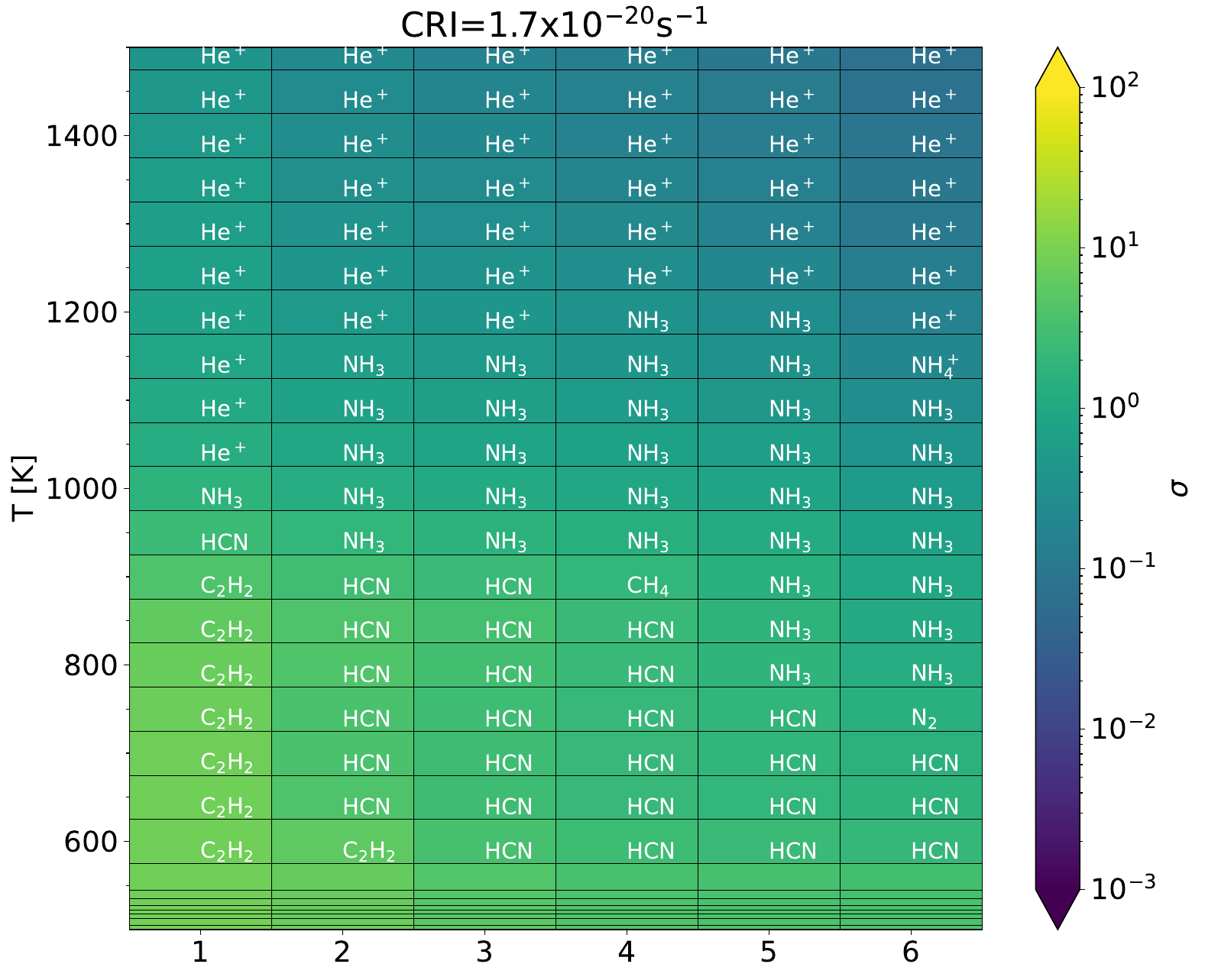}
& 
\includegraphics[width=0.5\textwidth]{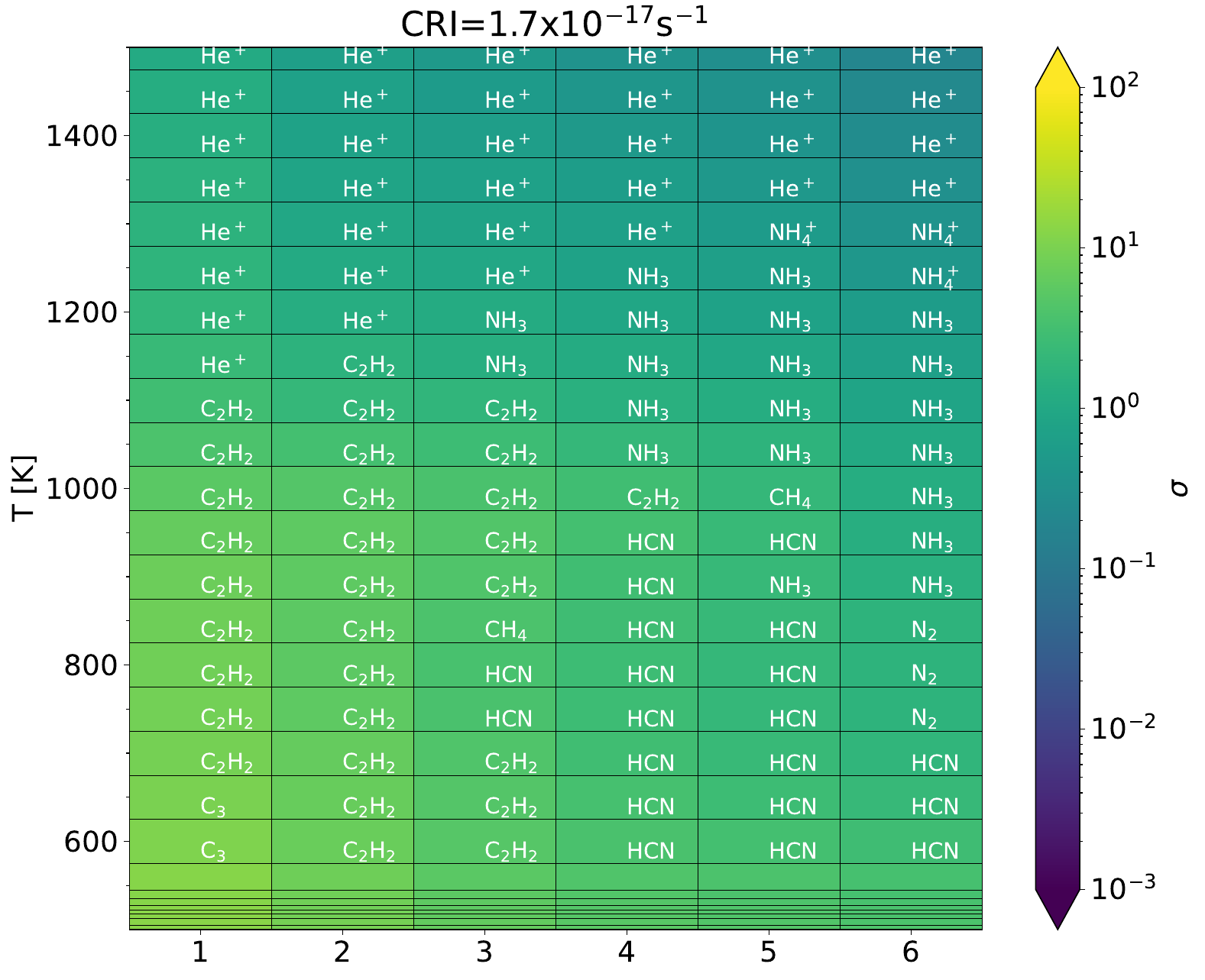}\\[-1mm]
{\# (density)} & {\# (density)}\\
\end{tabular}
\caption{Impact of cosmic-ray ionisation rate on deviations from thermodynamic equilibrium at different densities and temperatures from grid\_cr. The adopted values of CRIs are listed at the top of each panel. The x- and y-axes are the same as in Fig.\,\ref{without_photo}.} 
\label{CRI}
\end{figure*}

As shown in Fig.\,\ref{without_photo}, the deviations are $<\!1\%$ across all densities and temperatures considered, implying that our new {\sc ChaiTea} chemical network is capable of relaxing towards thermodynamic equilibrium, as expected. These deviations are caused by small differences in the thermochemical data between {\sc ProDiMo} and GGchem, and due to the different selections of molecular species. 

The species that contributes the most to $\sigma$ is marked by the white labels in Fig.~\ref{without_photo}. We do not observe any ions because these models are in thermodynamic equilibrium and the deviations are caused mostly by the abundant neutrals. At the lowest temperatures, there are a few cases where the {\sc ProDiMo} models do not converge, as indicated by the blank rectangles. The deviations are much larger there, only for numerical reasons.  In all the cases where the {\sc ProDiMo} models converge, the deviations from thermodynamic equilibrium are found to be small ($<\!1\%$). Hence, we consider our first test to be passed: the {\sc ChaiTea} network relaxes towards thermodynamic equilibrium down to $\sim$600\,K in the absence of any processes causing deviations from it.

\subsection{Effect of photoreactions}

We relaxed condition number (iv) (see Sect.\,\ref{Verification of the chemical network}), leading to grid\_photo, and perform a similar analysis to study the effect of photoreactions on thermodynamic equilibrium. As shown in Fig.\,\ref{with_photo}, the maximum deviation, $\sigma$, is then of the order of 10$^{-1}$. The radiative association reactions become relatively more significant at low densities, causing larger differences in the particle densities between GGchem and {\sc ProDiMo}. The deviations at higher densities are caused by ions such as \ce{H3O+} and \ce{H-}, at all temperatures. As the ions are more abundant in {\sc ProDiMo} relative to GGchem, this results in larger deviations from thermodynamic equilibrium.

There is a gradual increase in $\sigma$ going towards low densities. This is because the photoreactions scale with $\nH$. Higher temperatures are affected more as the rates of these photoreactions increase with temperature (Planck radiation field). Reactions such as 
\begin{eqnarray}
    \ce{H} + \ce{e-} \rightarrow \ce{H-} + h\nu
\end{eqnarray}
cause species such as \ce{e-} and \ce{H-} to deviate from detailed balance. \ce{H3O+} is not in perfect detailed balance due to 
\begin{equation}
    \ce{SiOH+} + \ce{H2O} \rightarrow \ce{H3O+} + \ce{SiO} \\
\end{equation}
which is an auto-generated reaction (see Sect.\,\ref{chemnetwork}).

\subsection{Effect of cosmic rays}

In this section, photo rates were again turned off. The deviations, $\sigma$, are presented in Fig.\,\ref{CRI} for various CRIs at different densities and temperatures. The deviations increase with increasing CRI. 

Across all densities and CRIs, $\sigma$ increases strongly with decreasing temperature. This is because the CRIs are independent of temperature, whereas the neutral-neutral gas-phase reaction rates decrease strongly with temperature because of the activation barriers. A secondary trend is that $\sigma$ increases for lower densities.  This effect is because the CRI scale with $\nH$, whereas the gas-phase chemical reactions scale with $\nH^2$ or $\nH^3$. 

As the CRI increases, the ions are affected first as they are the primary species created by the cosmic rays. Eventually, the neutrals are also affected at all temperatures and densities. High levels of cosmic rays create more reactive radicals such as \ce{OH}, and \ce{CH3}, which promote neutral-neutral chemistry. This results in neutrals deviating from thermodynamic equilibrium. For instance, at 1000\,K and a density of $\nH\!=7.562\times 10^{13}\rm\,cm^{-3}$, we observe that with increasing CRI, the molecule that causes the largest deviations shifts from \ce{He+} to \ce{NH3}. At a CRI of $1.7\times 10^{-17}\rm\,s^{-1}$, which is typically used in disk models, $\sigma$ is of the order of 1 at high temperatures (>1100\,K), but as large as 30 at low temperatures; i.e. the abundances in kinetic chemical equilibrium deviate from thermodynamic equilibrium by 30 orders of magnitude on average. 

\subsection{The chemical deviations in a full disk model}

In this section, we study the abundances in a 2D disk model around a T\,Tauri star, and evaluate the differences between the chemical solutions obtained with {\sc ProDiMo} in kinetic chemical equilibrium and with GGchem in thermodynamic equilibrium. The disk model is outlined in Sect.\,\ref{Disk modelling} and our choices of stellar, disk shape, dust size and opacity parameters are listed in Table\,\ref{input_parameters}.  We excluded ices, PAHs and excited \ce{H2}$^\star$ for this comparison, because these species are not included in GGchem. The model has a grid resolution of 200\,$\times$\,300 and includes UV, cosmic rays and X-ray chemistry. The second disk model, called GGchem, uses GGchem and the densities and gas temperatures calculated in the {\sc ProDiMo} model.

Figures \ref{abundances} and \ref{abun2} show the abundances of various molecules from these models, demonstrating how different these solutions are. As \cite{Woitke2021} discuss, the molecular abundances in thermodynamic equilibrium at low temperatures ($T\!<\!500\,$K) have a binary character: either a molecule is very abundant and is the major carrier of an element or it only attains a trace abundance. This is because at low temperatures the contribution of entropy term is low and the code maximises the binding energy of the molecules. In the case of the elemental abundances considered in this paper, these molecules are \ce{H2}, \ce{H2O}, \ce{CH4}, \ce{NH3}, \ce{H2S}, \ce{SiH4}, \ce{Mg(OH)2}, \ce{Fe(OH)2}, and \ce{(NaOH)2}. All other molecules have extremely low abundances in the GGchem model in the outer disk regions where the temperature is low. Molecules such as \ce{CO}, \ce{CO2}, and \ce{N2} are only present in the GGchem model in a relatively thin intermittent surface layer where the temperatures are $\sim$1000\,K due to the heating by UV processes. Temperatures exceed 5000\,K above this layer, which is too high for molecules to exist in high abundances.

\begin{figure}
   \centering
   {\sc ProDiMo} \hspace*{28mm} GGchem\\ 
   \includegraphics[width=\columnwidth,trim=10 5 10 5,clip]{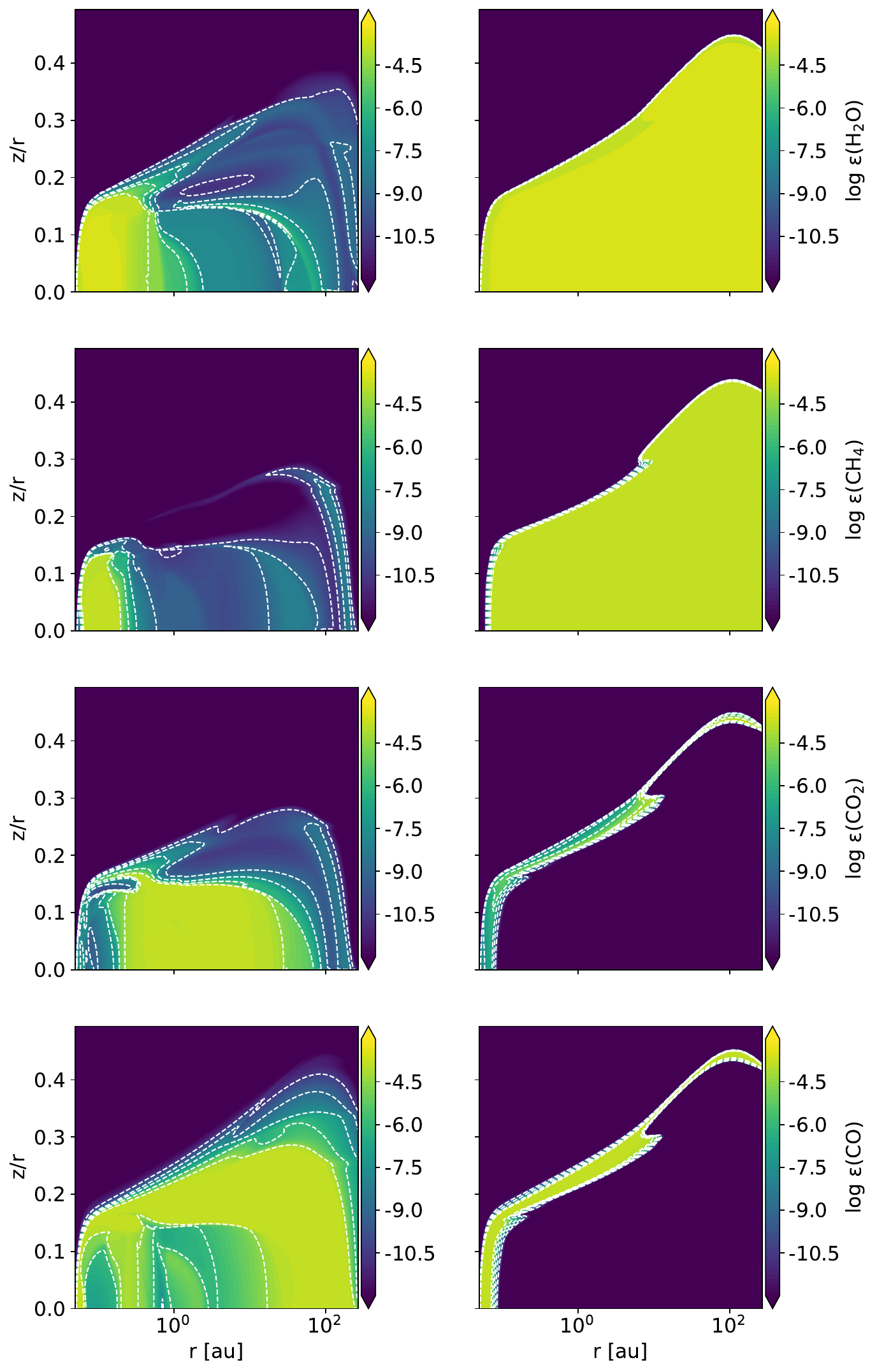}
   \vspace*{-5mm}
   \caption{Abundances of selected species in the 2D disk model calculated using chemical kinetics (left) and thermodynamic equilibrium (right). The white contours correspond to the tick values on the colorbars.}
   \vspace*{-2mm}
  \label{abundances}
\end{figure}
\begin{figure}
   \centering
   {\sc ProDiMo} \hspace*{28mm} GGchem\\ 
   \includegraphics[width=\columnwidth,trim=10 5 10 5,clip]{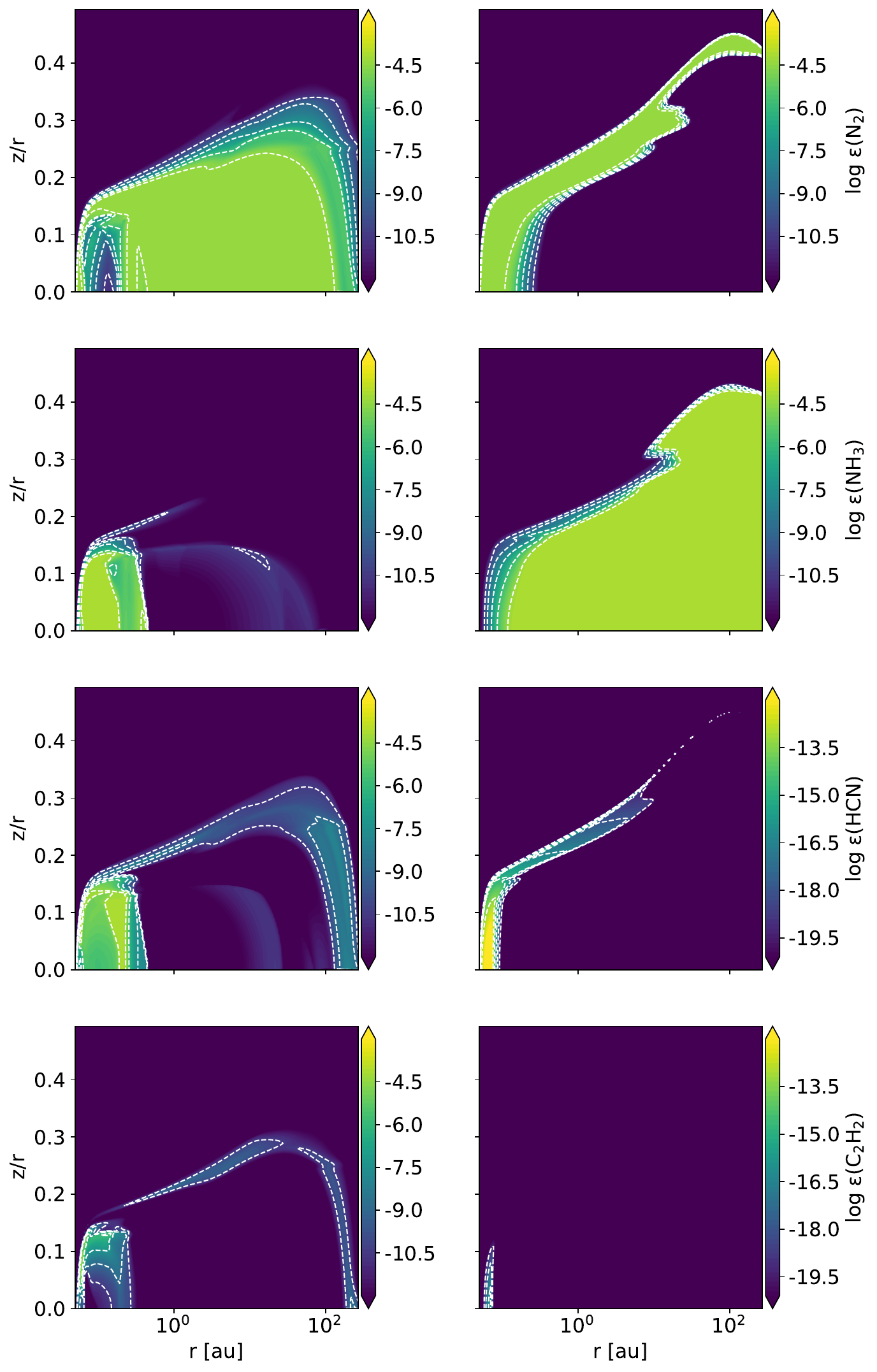}
   \vspace*{-5mm}
   \caption{Continued from Fig.~\ref{abundances}. We note the different scaling of the GGchem results for HCN and \ce{C2H2}.}
  \label{abun2}
\end{figure}

The abundances calculated with chemical kinetics by the {\sc ProDiMo} model are different (Figs\,\ref{abundances}, \ref{abun2}, left column). The cosmic rays constantly create small amounts of ions, atoms, and neutral radicals, that drive all the important gas-phase reactions and prevent the chemistry from relaxing towards thermodynamic equilibrium.

In particular, oxygen is stored in \ce{H2O} in thermodynamic equilibrium, whereas it is distributed among \ce{H2O}, \ce{CO2}, and \ce{CO} in kinetic equilibrium. Carbon is locked in \ce{CH4} in thermodynamic equilibrium throughout the cold disk, whereas it is found in the form of \ce{CO2} and \ce{CO} in kinetic equilibrium, similarly to in \cite{Bergin2009}. The main carrier of N in thermodynamic equilibrium is \ce{NH3} in the entire cold disk, but switches to \ce{N2} in the warm disk surface. In kinetic chemical equilibrium, N mostly forms \ce{N2} with \ce{HCN} and \ce{NH3} in the inner disk ($\la 0.5$\,au), as seen in Fig.\,\ref{abun2}.

\begin{figure}
   \centering
   \includegraphics[width=\linewidth]{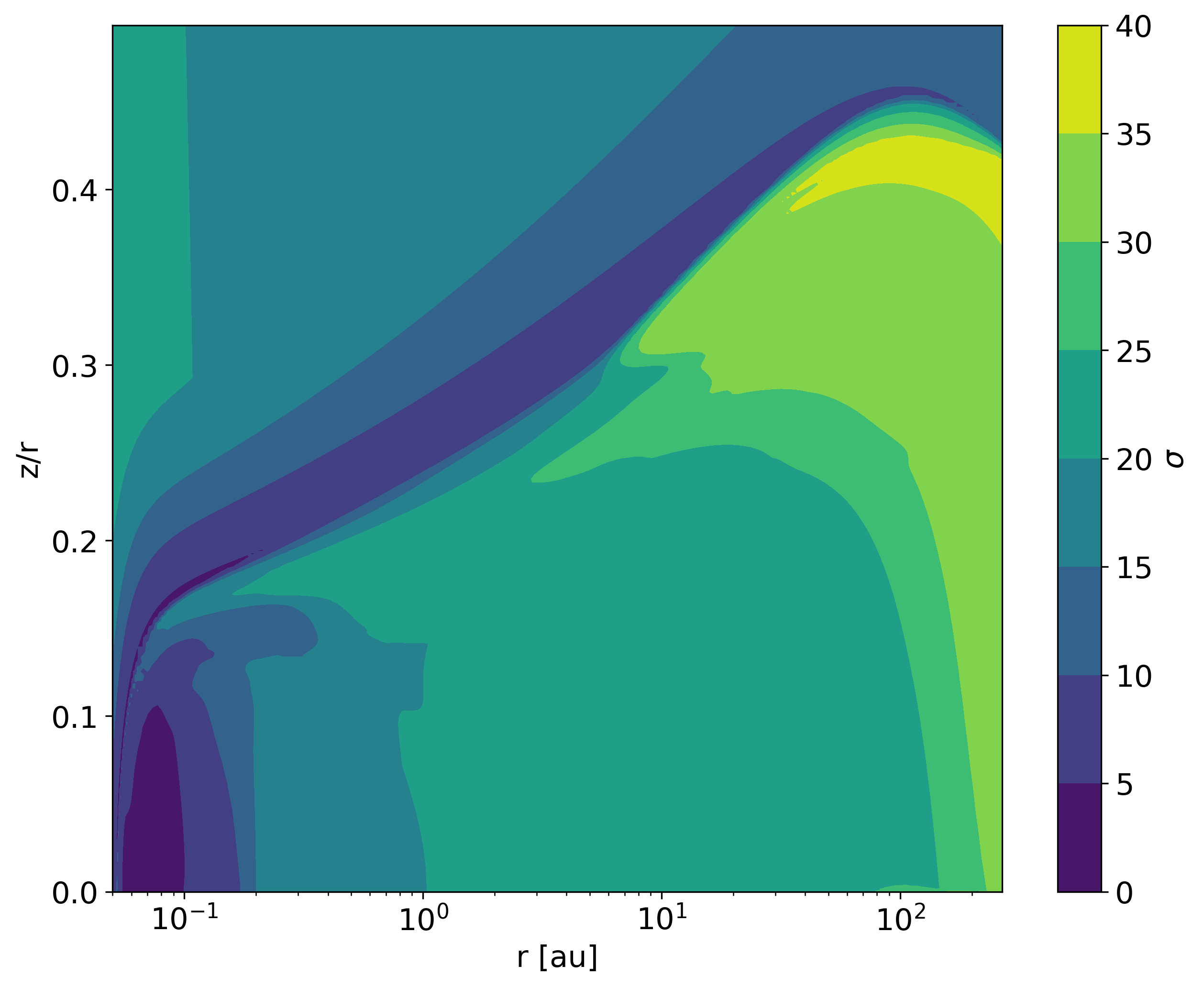}
   \caption{Deviation ($\sigma$) between kinetic chemical equilibrium and thermodynamic equilibrium in the 2D disk model. $\sigma\!=\!1$ denotes a deviation of one order of magnitude between the {\sc ProDiMo} and GGchem abundances.}
  \label{full_disk}
\end{figure}

Figure\,\ref{full_disk} shows the average chemical deviation, $\sigma$, between the two models as a function of disk radius, $r$, and height over the midplane, $z$. The minimum $\sigma$ is $\sim\!1$, indicating that thermodynamic equilibrium likely does not exist in any disk region. The inner midplane regions of the disk, directly behind the inner rim inside about 0.1\,au, are closest to thermodynamic equilibrium, because radiation such as cosmic rays is shielded here. As we go either vertically up or radially outwards in the disk, $\sigma$ becomes larger. These large $\sigma$ are due to the effect of irradiation by UV, X-rays, and cosmic rays. The outer midplane is affected by cosmic rays whereas UV and X-rays affect the chemistry in warm surface layers and prevent relaxation towards thermodynamic equilibrium. There is, however, the warm molecular intermittent surface layer where $\sigma$ again reaches values as low as 5.  Here, the gas-phase chemical reactions can compete with the cosmic-ray, X-ray, UV-induced chemistry, because temperatures are high enough to overcome the activation energies of the neutral-neutral reactions. However, the chemistry is still not in thermodynamic equilibrium. We also tested another criterion where $\sigma$ is only affected when $\big(\max\{n_i,\neqq_i\}/\nH\big)\,>\,10^{-13}$. However, we still observe a similar deviation between the kinetic chemical equilibrium and the thermodynamic equilibrium.

\section{Implications for disk kinetic chemistry} 
\label{Discussion_ch4}

\begin{table*}[!t]
    \centering
    \caption{Summary of chemical models used in Sect.~5.}
    \vspace*{-2mm}
    \label{tab:rev_models}
    \resizebox{\linewidth}{!}{%
    \begin{tabular}{cc|c|c|c|c} 
    \hline
    Model name & Databases used & \# species & reverse reac. & \# exothermic reac. & \# endothermic reac.\\ 
    \hline
    &&&&\\[-2.1ex]
    DIANA-standard & UMIST2022 $+$ DIANA reac. & 239 & $\times$ & 2117 & 158 \\
    DIANA-standard-rev & UMIST2022 $+$ DIANA reac. & 239 & $\checkmark$ & 2205 & 2138 \\
    {\sc ChaiTea} & UMIST2022 $+$ DIANA reac. $+$ STAND & 239 & $\checkmark$ & 2530 & 2464 \\ 
    \hline
    \end{tabular}}
    \tablefoot{The abbreviation "reac." means reactions.}
\end{table*}

\begin{figure*}[th]
   \centering
   \includegraphics[width=\linewidth]{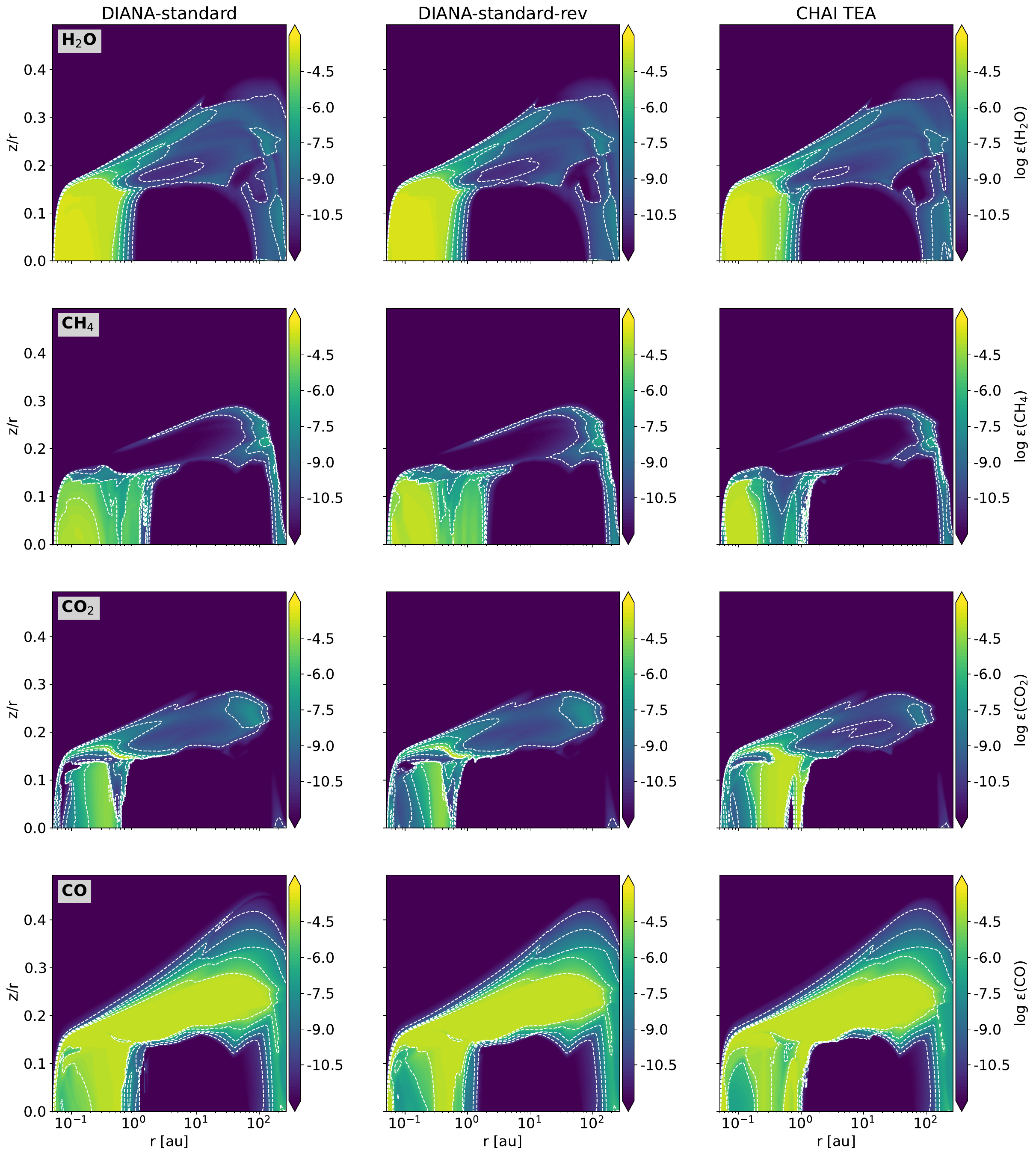}
   \caption{The abundances of various species in the three models calculated using different chemical networks and rate databases outlined in Table\,\ref{tab:rev_models}. The white dashed contours correspond to the tick values on the colourbar.}
  \label{abun_rev}
\end{figure*}

\begin{figure*}[th]
   \centering
   \includegraphics[width=\linewidth]{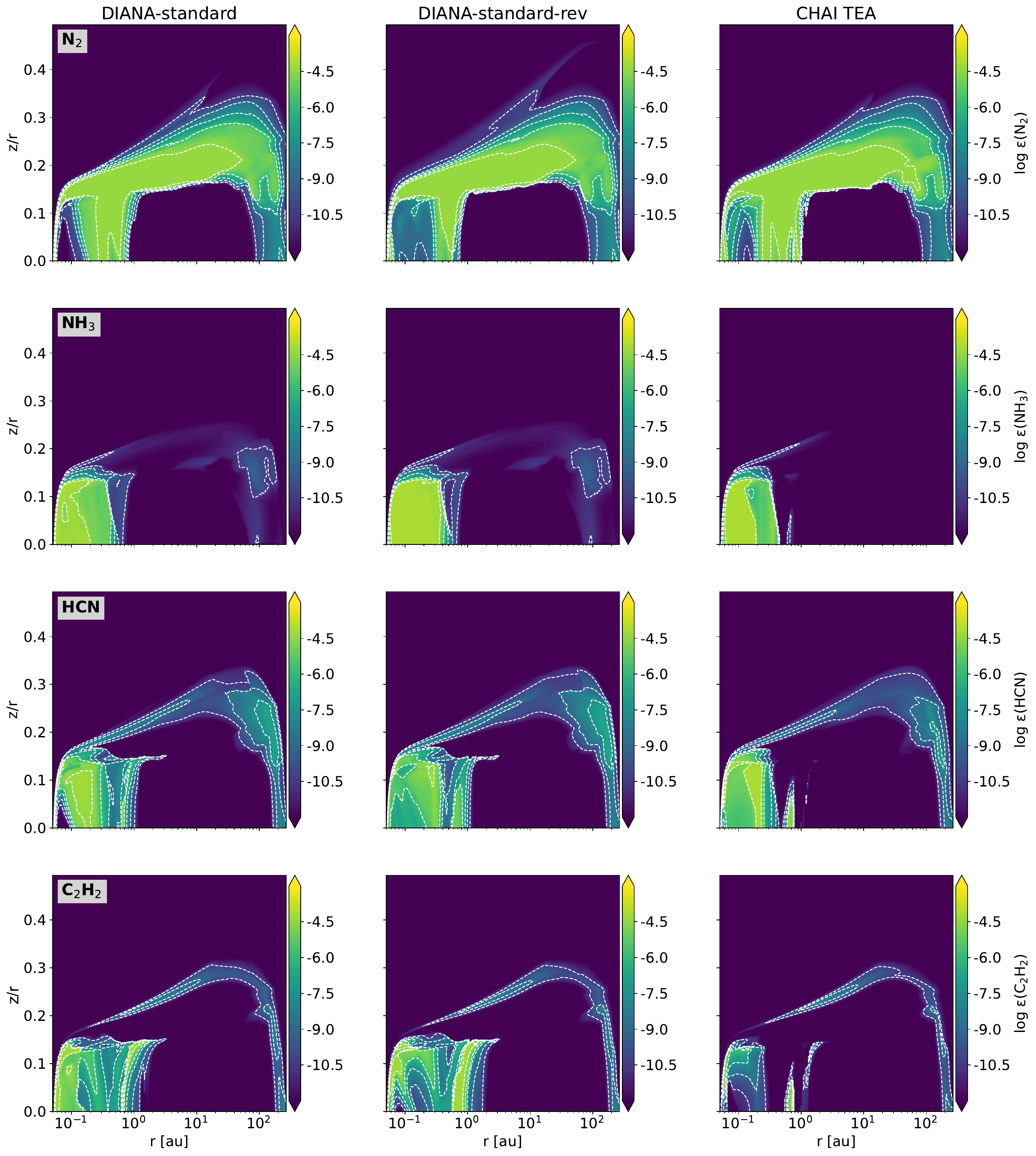}
   \caption{Continued from Fig.\,\ref{abun_rev}}
  \label{abun_rev1}
\end{figure*} 

To assess the implications of the above work for kinetic chemical networks, we studied the effect of extending the large DIANA standard network \citep{Kamp2017} to the new {\sc ChaiTea} chemical network on the 2D abundances in thermochemical disk models. We investigated this question in two steps; see Table\,\ref{tab:rev_models}.  First, we added the reverse reactions based on the same Gibbs free energies as discussed in this paper, and second, we added the STAND reactions which have preference over the UMIST 2022 and DIANA reactions. 

In all these models, the same 239 species were selected. We added excited molecular hydrogen H$_2^{\star}$, five charging states of PAHs, and the ices of all neutral species to the list of species in Table\,\ref{tab:species}.  Compared to the original large DIANA \citep{Kamp2017} standard network consisting of 235 species, there was also \ce{N2O}, \ce{N2O+}, \ce{HNO2+}, and \ce{N2O\#}, as we needed \ce{N2O} for the Si hydride reactions (Table~\ref{Si_table}). Following the rules laid out by \cite{Kamp2017} for developing a network, we also added \ce{N2O+} and \ce{HNO2+}. This was our final {\sc ChaiTea} network. The disk model was the same as in the previous section, see Table\,\ref{input_parameters}. The gas temperature was consistently calculated with the chemistry in each of the following models. 
UV-photo, X-ray and cosmic rays processes were included.

\subsection{Effect of adding the reverse reactions}

Table~\ref{tab:rev_models} shows that the inclusion of the reverse reactions mostly increases the number of endothermic reactions, as most of these are derived from their exothermic counterparts. It also "auto-cleans" the network by correcting spurious endothermic reactions with zero or extremely low activation energies, which are unrealistic, as discussed in Sect.~\ref{sec:reverse_rates}.

Our results are presented in Figs.~\ref{abun_rev}, \ref{abun_rev1}, and \ref{abun_rev_big}. The left columns in these figures show the model with the large DIANA and UMIST\,2022 reactions, whereas the middle column also includes the reverse of these reactions. By comparing the left and middle columns in these figures, we observe that the abundances of most molecules are only slightly affected (for example, \ce{H2O}). However, certain species show significant differences, particularly atomic H close to the outer midplane, and \ce{C2H2}, CO, \ce{CO2}, and \ce{N2} close to the inner midplane. For these species, the abundances decrease significantly when reverse reactions are included. Additionally, we observe an increase in \ce{N2H+} abundance in several disk regions when reverse reactions are considered. 

The relatively high abundance of atomic H in the outer midplane of the DIANA-standard model is an example of artefacts caused by spurious endothermic reactions with zero activation energy. One such reaction is
\begin{equation} 
\rm Ar + H_3^+ \to Ar^+ + H + H_2 \ . \quad\mbox{(erroneous)} 
\end{equation}
This reaction was included as one of the additional reactions collected by \cite{Kamp2017}, referencing \cite{Anicich1993}. However, there was an error in its inclusion; the correct reaction is
\begin{equation} 
\rm Ar + H_3^+ \to ArH^+ + H_2 \ . \quad\quad\mbox{(correct)} 
\end{equation}
The erroneous reaction has a reaction enthalpy of about +6.5 eV. While the correct reaction is also endothermic, with +0.43 eV, it is reported as barrier-free with zero activation energy (see \citealt{RainerJohnsen2005} for details). In the outer midplane, where C-, N-, O-, and S-bearing species are locked in their respective ices, \ce{H3+} is the most abundant molecular cation, maintaining charge balance with free electrons \citep{Balduin2023}. As a result, \ce{H3+} is relatively abundant, as is Ar, which leads the model with the erroneous reaction to produce a significant amount of neutral H in the midplane. However, when reverse reactions are included, the forward rate of this reaction is automatically corrected, as it would be for the correct reaction, and the artefact is eliminated. We identify 45 such spurious reactions in the large DIANA-standard network, and 68 when the STAND reactions are added, demonstrating once again that this is a common issue in astrochemistry \citep{Tinacci2023}.

We identify the following reasons for the increase in \ce{N2H+} abundance in various disk regions when reverse reactions are considered. First, the following constructed endothermic formation reactions become significant at very high temperatures (several thousand Kelvin):
\begin{eqnarray} 
\rm N + NH &\to& \rm N_2H^+ + e^-\\ 
\rm N_2 + H &\to& \rm N_2H^+ + e^- \ . 
\end{eqnarray}
Second, one of the dominant \ce{N2H+} destruction reactions:
\begin{equation} 
\rm N_2H^+ + O \to N_2 + OH^+ 
\end{equation}
is found to be endothermic ($+0.12$\,eV) but is reported as barrier free. This means that at temperatures as low as 20\,K, the rate of this reaction is automatically corrected to very low values, leading to an increase in the \ce{N2H+} abundance.

The abundance structures of \ce{H2O}, CO, and \ce{NH3} are similar between the two models. The abundance of \ce{c-C3H2} has three reservoirs. The innermost reservoir within 0.1\,au decreases in abundance when the reverse reactions are taken into account, whereas the reservoir around 1\,au increases. A similar trend is observed for \ce{CH3} at 1\,au. The inner reservoir of \ce{HCN} is smoother in abundance when including the reverse reactions. 

The abundance of \ce{C2H2} drops in the inner disk ($\la\!0.1\,$au) between DIANA-standard and DIANA-standard-rev. The reactions mostly contributing to the formation of \ce{C2H2} in DIANA-standard are 
\begin{eqnarray}
\ce{C2H3} + \ce{H} \rightarrow \ce{C2H2} + \ce{H2} \\
    \ce{C2H} + \ce{H2} \rightarrow \ce{C2H2} + \ce{H}.
\end{eqnarray}
These reactions are also reported as dominant by \cite{Kanwar2023} (grid point 3, fiducial model). However, when reverse reactions are included, each of these major formation reactions is balanced by the reverse of these reactions, thus resulting in no net contribution in the inner region (similar to detailed balance). Subsequently, the destruction of \ce{C2H2} by cosmic rays becomes dominant, leading to a decrease in abundance. These changes caused by the addition of reverse reactions can be crucial for the interpretation of JWST observations that can probe these deeper inner disk regions. 

\subsection{Effect of expanding to the {\sc ChaiTea} network}

When comparing the DIANA-standard-rev model to the {\sc ChaiTea} model (the middle and right columns of Figs.\,\ref{abun_rev}, \ref{abun_rev1}, \ref{abun_rev_big}), we observe that many molecular abundances are affected in various ways. This outcome is expected, as we replaced the UMIST 2022 reaction data with the STAND reaction set, leading to significant modifications in most of the important chemical reactions. However, the overall impact is not substantial. For example, the abundances of important species such as \ce{H2O}, CO, and \ce{e-} remain largely similar between the two models. In light of the mid-infrared and (sub-)millimetre observations, we discuss below a few species that show large differences in their abundance structures between the DIANA-standard-rev and {\sc ChaiTea} models in more detail. 

Between the two models, there are significant differences in the inner midplane that is probed by JWST. For example, the abundance of \ce{CO2} increases within 1\,au when using the {\sc ChaiTea} chemical network. This is due to more formation of \ce{CO2} in {\sc ChaiTea} relative to the DIANA-standard-rev model.
The dominant formation reaction in {\sc ChaiTea} model is 
\begin{equation}
    \ce{CO} + \ce{OH} \rightarrow \ce{CO2} + \ce{H}.
\end{equation}
This reaction is barrier free according to the STAND network. However, \cite{doi:10.1021/j100149a027} and \cite{Millar2024} report an activation barrier of 176 K. We investigated whether incorporating this barrier would alter the \ce{CO2} abundance structure, but find no significant changes. As a result, the final {\sc ChaiTea} network retains the reaction as barrier free, consistent with the STAND network. This can be further revisited in future work. The second notable change is a 2 orders of magnitude decrease in \ce{C2H2} abundance in the inner and central midplane when using the {\sc ChaiTea} network. \cite{Kanwar2023} also report a reduction in \ce{C2H2} abundance in the inner midplane when larger hydrocarbons were included in an extended hydrocarbon chemical network. These larger hydrocarbons are not included in the current network and are beyond the scope of this study. Integrating the two networks could thus lead to substantially lower \ce{C2H2} abundances but eventually provide a more accurate abundance structure for mid-infrared observations \citep{Tabone2023, Arabhavi2023, Xie2023, Kanwar2024, Kamber2024}. The change in HCN abundance is more complex. In the DIANA-standard-rev model, the thin, higher-abundance surface layer extends farther radially than in the {\sc ChaiTea} model (see Fig\,\ref{abun_rev1}). There is less formation of \ce{HCN} in the latter. The major formation pathway in the surface layer in both the models is
\begin{equation}
    \ce{CN} + \ce{H2} \rightarrow \ce{HCN} + \ce{H}.
\end{equation}
In the DIANA-standard-rev model, this reaction has a very low energy barrier (6.3$\cdot$10$^{-4}$\,K). However, in the {\sc ChaiTea} model, the reaction has a higher barrier of 2.3$\cdot$10$^3$\,K, resulting in a decrease in \ce{HCN} abundance in the surface layer by an order of magnitude. The abundance of some ions such as \ce{CH+} increases in the surface layers of the outer disk in the {\sc ChaiTea} model and this may affect the interpretation of the Herschel Space Telescope observations \citep{Thi2011}.

There are also significant differences in the outer disk that may affect the inferences from ALMA observations. For example, the abundance of HCN decreases in the {\sc ChaiTea} model in the surface layers of the outer disk relative to DIANA-standard-rev by at least 1.5 orders of magnitude. This could lead to lower line fluxes at millimetre wavelengths when using the {\sc ChaiTea} chemical network.
Other neutral molecules such as \ce{NH3}, \ce{C2H}, \ce{CH3}, and \ce{c-C3H2} also show substantially lower abundances in the outer disk when using the {\sc ChaiTea} chemical network. \ce{C2H} plays a key role in determining the carbon budget in the outer disk. The abundance of the key molecular ion \ce{N2H+} on the other hand increases in the outer disk when using the {\sc ChaiTea} chemical network. \cite{Qi2015} and, \cite{Vanthoff2017} used \ce{N2H+} to infer the location of the CO iceline in the disk as it is difficult to observe directly. There is also a decrease in the abundances of \ce{CH3} and \ce{c-C3H2} in the inner disk by 2-3 orders of magnitude in the {\sc ChaiTea} model compared to DIANA-standard-rev. This can help to interpret JWST \citep{Arabhavi2023, Kanwar2024} and ALMA \citep{Qi2013a, Bergin2016} observations better.

\section{Summary and conclusions}\label{Conclusion_ch4}

We merged the STAND chemical network developed for planetary and exoplanetary atmospheres \citep{Rimmer2016, Rimmererr2019}, which includes pressure-dependent termolecular and reverse reactions based on Gibbs free energy data, with the UMIST~2022 chemical network and our large DIANA standard chemical network developed for disks \citep{Kamp2017}, which was later updated in \cite{Kanwar2023}. A cornerstone in developing the new network was compiling a comprehensive set of Gibbs free energy data from various sources. For some molecules, data had to be estimated using ionisation potentials and proton affinities to ensure all gas-phase reactions could be reversed. This new, integrated chemical network is named {\sc ChaiTea}.

We then studied under what conditions thermodynamic equilibrium can be established in planet-forming disks. Finally, we explored how the chemical composition of the disk changed when we used the new {\sc ChaiTea} chemical network instead of the large DIANA standard network. We used a 2D T\,Tauri disk model for this investigation, including UV, X-ray, and cosmic-ray chemistry. 

\begin{itemize}
    \item In the absence of any UV, X-ray or cosmic-ray irradiation, we verify that the new {\sc ChaiTea} network converges towards thermodynamic equilibrium by benchmarking it against the chemical equilibrium code GGchem \citep{Woitke2018}.
    \item However, at a gas density of 7.56$\times$10$^{13}$cm$^{-3}$ ($\sim$10\,dyn/cm$^2$), the chemical relaxation timescale $\tau_{\rm chem,3}$, towards thermodynamic equilibrium can be as short as a month at 2000\,K, which increases to $\sim$$10^8$\,yr at 1200\,K, and exceeds the age of the universe at 1100\,K.
    \item The chemical relaxation proceeds simultaneously via fast and slow chemical modes with characteristic timescales that differ by many tens of orders of magnitude. We find that one of the slowest modes is the production of \ce{N2} from other N-bearing molecules after the neutral atoms are consumed.
    \item The inclusion of radiative association, direct recombination reactions (considered non-reversible), and photo rates based on a local Planck radiation field, $J_\nu\!=\!B_\nu(T)$, causes deviations from thermodynamic equilibrium by 2-20\% across all densities and temperatures studied.
    \item The inclusion of cosmic rays leads to significant deviations from thermodynamic equilibrium, initially affecting the charged molecules and eventually impacting the abundant neutral species. At low temperatures ($T\!\la\!750\,$K), even a tiny CRI of $10^{-40}\,\rm s^{-1}$ alters the neutral chemistry. At the standard CRI of $1.7\times 10^{-17}\,\rm s^{-1}$, below $\sim$1200\,K, deviations from thermodynamic equilibrium can exceed 10–30 orders of magnitude.
    
    \item In a 2D disk model, no region is found to be in thermodynamic equilibrium.  We find that a small, warm, dense region in the midplane directly behind the inner rim ($r\!\la\!0.1\,$au), where the cosmic rays are mostly shielded, is closest to thermodynamic equilibrium, with $\sigma \sim$\,1. There is a warm intermittent molecular layer where $\sigma \sim$\,5. In all other regions $\sigma$ is much larger. 
    \item In kinetic equilibrium, nitrogen is primarily stored in \ce{N2}, carbon in \ce{CO} and \ce{CO2}, and oxygen in \ce{H2O} and \ce{CO}, whereas the thermodynamically favoured molecules at low temperatures are \ce{NH3} for nitrogen, \ce{CH4} for carbon, and \ce{H2O} for oxygen. 
    
    \item When using the {\sc ChaiTea} chemical network, the change in the abundances of species such as \ce{CO}, \ce{H2O} and free electrons is not major (an order of magnitude). The abundance of species such as \ce{C2H2}, \ce{HCN} and \ce{CH4} decreases in the inner midplane, while the abundance of \ce{CO2} increases within 1\,au. Some molecular cations such as \ce{N2H+} and \ce{CH+} attain larger abundances in the intermittent warm molecular layer and in the outer disk regions. The reasons for these differences are complex. Differences in the rate coefficients of the leading reactions can have profound effects on the resulting abundances. Defining reaction pairs based on Gibbs free energy allows for the identification and automatic correction of spurious endothermic reactions reported with zero or minimal activation energies, which are chemically implausible. Such reactions can significantly impact the chemistry of the outer disk.
\end{itemize}
The techniques implemented in this work, and the new chemical network developed, can be used to interpret both the (sub-) millimetre molecular lines observable with ALMA and the mid-infrared observations with JWST. The latter probes the inner, dense, and warm regions where this network is more complete. The {\sc ChaiTea} chemical network can also be used to quantify the significance of the termolecular and reverse reactions using the observations. 

\begin{acknowledgements}
This project has received funding from the European Union’s Horizon 2020 research and innovation programme under the Marie Sklodowska-Curie grant agreement No. 860470. We thank the anonymous referee for their constructive comments. We acknowledge discussions with W.F. Thi. J.K. acknowledges productive discussions with Till Kaeufer, Nidhi Bangera, David Lewis, and Ashwani Rajan.
\end{acknowledgements}
\bibliographystyle{aa}
\bibliography{Paper}
\appendix
\twocolumn
\section{Derivation of the law of mass action}
\label{LawMassAction}

We assume a mixture of ideal gases with partial pressures $p_i =N_i\,RT/V$, where $N_i$ are the mol numbers of the gas species $i$, $V$ is the common volume, and $R\rm\,[J/K/mol]$ is the ideal gas constant. $P\!=\!\sum_i p_i$ is the total gas pressure. $(\Tst,\pst)$ denotes a thermodynamical reference state at room temperature $\Tst\!=\!298$\,K and at standard pressure $\pst\!=\!1$\,bar, to which some of the thermodynamic quantities refer to. The thermodynamic description of such mixtures, see \citet{Atkin2006}, is
\begin{eqnarray}
  G &=& U + PV - TS  \\
  U &=& \sum_i N_i \left(\ust_i + \!\!\int_\Tst^T\!\!c_{V,i}\;dT \right) \\
  PV &=& \sum_i N_i\,RT \\
  S &=& \sum_i N_i\,\Bigg(\underbrace{\sst_i + \!\!\int_{\Tst,\pst}^{T,\pst} \!\!\Big(c_{V,i}+R\Big)\,\frac{dT}{T}}_{\displaystyle s_i(T,\pst)}\;-\underbrace{\int_{T,\pst}^{T,p_i} \!\!R\;\frac{dP}{P}}_{\displaystyle R\ln\frac{p_i}{\pst}} \Bigg)\ ,
  \label{entropy}
\end{eqnarray}
where $G$ is the system Gibbs free energy, $U$ the internal energy, and $S$ the entropy. The small symbols refer to the molar properties of the pure gases $i$: $\ust_i$ is the internal energy in the standard state $(\Tst,\pst)$, $c_{V,i}$ the heat capacity, and $\sst_i$ is the entropy in the standard state $(\Tst,\pst)$. $s_i(T,\pst)$ is the entropy of the pure gas $i$ at temperature $T$ and standard pressure $\pst$. The terms in the integrals in Eq.\,(\ref{entropy}) can be derived from the first law of thermodynamics in the form $dU=-PdV+TdS$ for processes of an ideal gas at constant temperature or pressure. The term $R\ln(p_i/\pst)$ is the mixing entropy. Altogether, the system Gibbs free energy is
\begin{equation}
  G = \sum_i N_i\,\Bigg( \underbrace{\ust_i + \!\!\int_\Tst^T\!\!c_{V,i}\;dT + RT - T s_i(T,\pst)}_{\displaystyle g_i(T,\pst)} ~+~ RT\ln \frac{p_i}{\pst} \Bigg) \nonumber \ ,
\end{equation}
where $g_i(T,\pst)$ is the molar Gibbs free energy of the pure gas $i$ at temperature $T$ and standard pressure $\pst$.

Next, we consider a hypothetical chemical reaction $\rm N_2 + 3\,H_2 \to 2\,NH_3$, where $k\!=\!3$, $\rm\{1\!\to\!N_2 , 2\!\to\!H_2 , 3\!\to\!NH_3\}$ and stoichiometric coefficients $\{\nu_1\!=\!-1 , \nu_2\!=\!-3 , \nu_3\!=\!+2\}$. We also consider a small number of mols of such reactions $\Delta$ to take place at constant temperature and pressure. The principle of minimisation of the system Gibbs free energy in this case requires that
\begin{equation}
  \Delta G ~=~ G(T,\Neq_1+\nu_1\Delta, ..., \Neq_k+\nu_k\Delta) - G(T,\Neq_1, ..., \Neq_k) ~=~ 0
  \label{thermoEq}
\end{equation}
in the thermodynamical equilibrium state that is characterised by the mol numbers $\Neq_i$.  From Eq.\,(\ref{thermoEq}) it follows that
\begin{equation}
  \Delta G ~=~ \Delta \sum_i \nu_i \,\bigg(g_i(T,\pst) + RT\ln \frac{\peq_i}{\pst} \bigg) ~=~0
  \label{MWG1}
\end{equation}
for any small $\Delta$. The Gibbs free energy data for the pure gases $i$ can be found in various thermochemical databases, here we use \cite{Burcat}. These data are given as the Gibbs free energy of formation $\dG_{\!\!i}(T)\rm\,[J/mol]$ of a molecule from the standard states of the elements at room temperature.  Since any chemical reaction obeys the principle of element conservation, these constants cancel in Eq.\,(\ref{MWG1}), so we can write this equation also as
\begin{equation}
  \sum_i \nu_i\,\dG_{\!\!i} ~=~ -\,RT\sum_i \nu_i \ln \frac{\peq_i}{\pst} 
  \label{MWG2} \ .
\end{equation}
Introducing the reaction Gibbs free energy as
\begin{equation}
  \dGR = \sum\limits_{i=1}^k \nu_i\;\dG_{\!\!i}
\end{equation}
we find the law of mass action
\begin{equation}
\prod_{i=1}^{k}\left(\frac{\peq_i}{\pst}\right)^{\nu_i}
     ~=~ \exp\left(-\frac{\dGR}{RT}\right) \ .
   \label{MWG3}  
\end{equation}
Equation (\ref{MWG3}) is valid for all hypothetical chemical reactions in thermodynamical equilibrium, even if they cannot proceed kinetically. We note that $\Delta G\!=\!0$ in thermodynamical equilibrium, whereas $\dGR\!\neq\!0$ because of the mixing entropy. In the main text in Sect.~\ref{backward} we consider the reaction $\rm A+B \to C+D+E$, where $k=5$ and $\{\nu_1\!=\!-1 , \nu_2\!=\!-1 , \nu_3\!=\!+1 , \nu_4\!=\!+1 , \nu_5\!=\!+1\}$.


\section{Thermochemical data of missing species in Burcat format}

Table~\ref{tab:newBURCAT} summarises our efforts to collect, convert, estimate or construct the thermochemical in Burcat format (NASA 7-polynomials) of those species missing in \cite{Burcat}.

\begin{table*}
\caption{Additional thermochemical data not included in \cite{Burcat}. $\dH(0\rm K)$ is given in kJ/mol.}
\label{tab:newBURCAT}
\vspace*{-3mm}
\resizebox{17.5cm}{!}{
\begin{tabular}{c|c|rrrrrrr|c|c}
\hline
&&&&&&&&&&\\[-2.1ex]
 & $\dH(0\rm K)$ & \multicolumn{7}{l|}{NASA 7-polynomial coefficients} & source & Ref.\\
&&&&&&&&&&\\[-2.1ex]
\hline
&&&&&&&&&&\\[-2.1ex]
\ce{SiH+ } & 1146.2 & 3.7397(+00) & 7.8649(-04) &-5.0816(-07) & 1.8261(-10) &-2.1808(-14) & 1.3663(+05) & 4.8601(-01) & GGchem & (1) \\
 & & 3.8119(+00) & 5.6816(-04) &-2.3534(-06) & 3.8922(-11) & 2.8413(-12) & 1.3686(+05) & 5.7500(-01) & & \\
\ce{SiO+ } & 1012.2 & 4.2399(+00) & 8.1021(-04) &-6.9523(-07) & 2.7118(-10) &-3.6990(-14) & 1.1977(+05) & 6.1025(-01) & GGchem & (2) \\
 & & 4.4826(+00) &-7.2246(-03) & 1.1014(-05) & 3.3442(-09) &-8.9184(-12) & 1.2068(+05) & 2.3318(+00) & & \\
\ce{SiH  } & 374.9 & 3.9266(+00) & 7.2061(-04) &-5.6154(-07) & 2.1249(-10) &-2.5291(-14) & 4.3788(+04) & 7.0652(-01) & GGchem & (1) \\
 & & 4.0128(+00) &-1.1455(-04) &-1.9727(-06) & 2.0099(-09) & 7.7663(-13) & 4.4090(+04) & 9.6061(-01) & & \\
\ce{SiH2 } & 256.3 & 4.1455(+00) & 2.5229(-03) &-5.2877(-07) &-3.5792(-11) & 1.4343(-14) & 3.0656(+04) &-1.8131(+00) & GGchem & (1) \\
 & & 3.9131(+00) & 2.1045(-02) &-3.4354(-05) &-1.7474(-08) & 4.1820(-11) & 2.9408(+04) &-5.9422(+00) & & \\
\ce{SiO  } & -103.0 & 4.2327(+00) & 6.9584(-04) &-6.8842(-07) & 2.5568(-10) &-2.9617(-14) &-1.3716(+04) & 5.1913(-01) & GGchem & (1) \\
 & & 4.3199(+00) & 1.4703(-03) &-5.5197(-06) &-5.1887(-10) & 6.2555(-12) &-1.3451(+04) & 4.3637(-01) & & \\
\ce{SiS  } & 103.4 & 4.4492(+00) & 6.9759(-04) &-8.2768(-07) & 3.1546(-10) &-3.6221(-14) & 1.1098(+04) & 8.7450(-01) & GGchem & (1) \\
 & & 4.5105(+00) & 1.8334(-03) &-4.9706(-06) &-1.1222(-09) & 5.2802(-12) & 1.1289(+04) & 6.7797(-01) & & \\
\ce{SiN  } & 370.7 & 4.2963(+00) & 8.7486(-04) &-7.1119(-07) & 2.7522(-10) &-3.2824(-14) & 4.3000(+04) & 5.5546(-01) & GGchem & (1) \\
 & & 4.4628(+00) &-2.8399(-03) & 2.3260(-06) & 5.7339(-09) &-5.8556(-12) & 4.3526(+04) & 1.2827(+00) & & \\
\ce{HS+  } & 1152.3 & 3.8260(+00) & 6.2009(-04) &-5.6669(-07) & 2.3006(-10) &-2.7417(-14) & 1.3714(+05) & 7.3004(-01) & GGchem & (2) \\
 & & 3.9039(+00) &-2.6695(-03) & 4.9694(-06) & 2.8348(-09) &-6.6726(-12) & 1.3748(+05) & 1.4508(+00) & & \\
\ce{SO+  } & 1007.0 & 4.3648(+00) & 5.5455(-04) &-6.0510(-07) & 2.4293(-10) &-3.0644(-14) & 1.1944(+05) & 1.0572(+00) & GGchem & (2) \\
 & & 4.4932(+00) &-3.8229(-03) & 5.7056(-06) & 4.1887(-09) &-7.8562(-12) & 1.1994(+05) & 1.9966(+00) & & \\
&&&&&&&&&&\\[-2.1ex]
\hline
&&&&&&&&&&\\[-2.1ex]
 \ce{SiH2+} & 1161.9 & 3.9586(+00) & 2.5888(-03) &-4.7539(-07) &-6.5668(-11) & 1.7826(-14) & 1.3965(+05) &-2.0336(+00) & $E_{\rm corr}$ & (3)\\
            &        & 3.7122(+00) & 2.1728(-02) &-3.4735(-05) &-1.9445(-08) & 4.3884(-11) & 1.3833(+05) &-6.3278(+00) &&\\
 \ce{SiH3+} &  998.9 & 5.9001(+00) & 3.9686(-03) &-1.5027(-06) & 2.5108(-10) &-1.5389(-14) & 1.1743(+05) &-8.0346(+00) & $E_{\rm corr}$ & (3)\\
            &        & 2.0650(+00) & 1.8174(-02) &-2.6340(-05) & 2.2832(-08) &-8.1296(-12) & 1.1842(+05) & 1.1172(+01) &&\\ 
 \ce{SiH5+} &  924.0 & 6.5037(+00) & 8.9121(-03) &-3.3100(-06) & 5.4608(-10) &-3.3178(-14) & 1.0703(+05) &-1.6332(+01) & $E_{\rm corr}$ & (3)\\
            &        & 8.4057(-01) & 2.4498(-02) &-1.9429(-05) & 7.6657(-09) &-9.5662(-13) & 1.0867(+05) & 1.3119(+01) &&\\
 \ce{SiOH+} &  655.6 & 4.2646(+00) & 2.5640(-03) &-6.0227(-07) &-2.2472(-11) & 1.3501(-14) & 7.8735(+04) &-2.2210(+00) & $E_{\rm corr}$ & (3)\\
            &        & 4.0193(+00) & 2.3313(-02) &-3.8282(-05) &-2.1974(-08) & 4.9363(-11) & 7.7380(+04) &-6.8520(+00) &&\\
 \ce{HSiS+} & 1014.3 & 4.2623(+00) & 7.6347(-04) &-7.7431(-07) & 2.8559(-10) &-3.2738(-14) & 1.2033(+05) & 6.5399(-01) & $E_{\rm corr}$ & (3)\\
            &        & 4.3096(+00) & 2.5161(-03) &-5.3514(-06) &-3.0933(-09) & 7.3449(-12) & 1.2045(+05) & 2.9236(-01) &&\\
 \ce{HSO2+} &  603.8 & 5.1973(+00) & 1.7452(-03) &-5.7868(-07) & 7.8590(-11) &-3.1856(-15) & 7.0054(+04) &-2.0518(+00) & $E_{\rm corr}$ & (3)\\
            &        & 3.4739(+00) & 2.9657(-03) & 8.0882(-06) &-1.5627(-08) & 7.8274(-12) & 7.0645(+04) & 7.5831(+00) &&\\
 \ce{ H3S+} &  803.4 & 2.7918(+00) & 3.6635(-03) &-1.1747(-06) & 1.6696(-10) &-8.1883(-15) & 9.4966(+04) & 6.5587(+00) & $E_{\rm corr}$ & (3)\\
            &        & 3.9193(+00) &-1.1964(-03) & 7.8335(-06) &-9.0353(-09) & 4.2070(-12) & 9.4729(+04) & 1.1461(+00) &&\\
 \ce{ HN2+} & 1040.5 & 5.4526(+00) & 1.3969(-03) &-4.9263(-07) & 7.8601(-11) &-4.6076(-15) & 1.2371(+05) & 4.7312(+00) & $\rm PA=493.8\,kJ/mol$ & (4)\\
            &        & 6.0310(+00) &-1.2366(-04) &-5.0300(-07) & 2.4353(-09) &-1.4088(-12) & 1.2358(+05) & 1.8268(+00) &&\\
 \ce{ H2S+} &  996.1 & 4.7879(-01) & 3.5976(-03) &-1.2280(-06) & 1.9683(-10) &-1.1672(-14) & 1.1858(+05) & 1.8500(+01) & $\rm IP=10.457\,eV$ & (4)\\
            &        & 1.6202(+00) &-1.8791(-03) & 8.2143(-06) &-7.0643(-09) & 2.1423(-12) & 1.1841(+05) & 1.3253(+01) &&\\
 \ce{ SiS+} & 1124.6 & 1.9492(+00) & 6.9759(-04) &-8.2768(-07) & 3.1546(-10) &-3.6221(-14) & 1.3404(+05) & 1.2595(+01) & $\rm IP=10.53\,eV$ & (4)\\
            &        & 2.0105(+00) & 1.8334(-03) &-4.9706(-06) &-1.1222(-09) & 5.2802(-12) & 1.3423(+05) & 1.2399(+01) &&\\
 \ce{HN2O+} & 1069.2 & 7.3231(+00) & 2.6270(-03) &-9.5851(-07) & 1.6001(-10) &-9.7752(-15) & 1.2597(+05) &-3.3424(+00) & $\rm PA=549.8\,kJ/mol$ & (4)\\
            &        & 4.7572(+00) & 1.1305(-02) &-1.3671(-05) & 9.6820(-09) &-2.9307(-12) & 1.2664(+05) & 9.6173(+00) &&\\
 \ce{  CS+} & 1374.9 & 1.2696(+00) & 7.3098(-04) &-2.4292(-07) & 2.8807(-11) &-5.2196(-17) & 1.6447(+05) & 1.5141(+01) & $\rm IP=11.33\,eV$ & (4)\\
            &        & 1.2312(+00) &-3.0980(-03) & 1.2483(-05) &-1.4163(-08) & 5.3337(-12) & 1.6467(+05) & 1.6269(+01) &&\\
 \ce{ HCS+} & 1019.4 & 6.2696(+00) & 7.3098(-04) &-2.4292(-07) & 2.8807(-11) &-5.2196(-17) & 1.2108(+05) & 2.2796(+00) & $\rm PA=791.5\,kJ/mol$ & (4)\\
            &        & 6.2312(+00) &-3.0980(-03) & 1.2483(-05) &-1.4163(-08) & 5.3337(-12) & 1.2127(+05) & 3.4079(+00) &&\\
 \ce{H2CS+} & 1028.7 & 1.6980(+00) & 5.1411(-03) &-1.9040(-06) & 3.3356(-10) &-2.1438(-14) & 1.2172(+05) & 1.3616(+01) & $\rm IP=9.38\,eV$ & (4)\\
            &        & 1.4889(+00) &-4.4809(-03) & 3.2315(-05) &-3.9856(-08) & 1.5780(-11) & 1.2222(+05) & 1.7019(+01) &&\\
 \ce{H3CS+} &  893.1 & 6.6980(+00) & 5.1411(-03) &-1.9040(-06) & 3.3356(-10) &-2.1438(-14) & 1.0477(+05) & 7.5474(-01) & $\rm PA=759.7\,kJ/mol$ & (4)\\
            &        & 6.4889(+00) &-4.4809(-03) & 3.2315(-05) &-3.9856(-08) & 1.5780(-11) & 1.0527(+05) & 4.1579(+00) &&\\
 \ce{ OCS+} &  934.1 & 2.8746(+00) & 2.1041(-03) &-7.7642(-07) & 1.2975(-10) &-7.9241(-15) & 1.1157(+05) & 7.9361(+00) & $\rm IP=11.18\,eV$ & (4)\\
            &        &-7.2801(-01) & 1.7149(-02) &-2.7308(-05) & 2.2555(-08) &-7.3437(-12) & 1.1235(+05) & 2.5402(+01) &&\\
 \ce{HOCS+} &  765.0 & 7.8746(+00) & 2.1041(-03) &-7.7642(-07) & 1.2975(-10) &-7.9241(-15) & 8.9513(+04) &-4.9254(+00) & $\rm PA=628.5\,kJ/mol$ & (4)\\
            &        & 4.2720(+00) & 1.7149(-02) &-2.7308(-05) & 2.2555(-08) &-7.3437(-12) & 9.0298(+04) & 1.2540(+01) &&\\
 \ce{  NS+} & 1139.8 & 1.3827(+00) & 6.6921(-04) &-2.5436(-07) & 4.3769(-11) &-2.4782(-15) & 1.3598(+05) & 1.5963(+01) & $\rm IP=8.87\,eV$ & (4)\\
            &        & 2.1131(+00) &-6.9704(-03) & 2.0142(-05) &-2.1237(-08) & 7.7909(-12) & 1.3604(+05) & 1.3504(+01) &&\\
 \ce{ HNS+} & 1080.9 & 6.3827(+00) & 6.6921(-04) &-2.5436(-07) & 4.3769(-11) &-2.4782(-15) & 1.2825(+05) & 3.1014(+00) & $\rm PA=7.59\,eV$ & (4)\\
            &        & 7.1131(+00) &-6.9704(-03) & 2.0142(-05) &-2.1237(-08) & 7.7909(-12) & 1.2830(+05) & 6.4288(-01) &&\\
 \ce{ SiC+} & 1615.1 & 1.1057(+00) & 1.0900(-03) &-2.6557(-07) & 2.8396(-11) &-1.1470(-15) & 1.9360(+05) & 1.8014(+01) & $\rm IP=9.1\,eV$ & (4)\\
            &        & 1.1187(+00) &-1.6984(-03) & 1.0545(-05) &-1.3659(-08) & 5.6517(-12) & 1.9367(+05) & 1.8457(+01) &&\\
 \ce{HCSi+} & 1590.9 & 6.1057(+00) & 1.0900(-03) &-2.6557(-07) & 2.8396(-11) &-1.1470(-15) & 1.9004(+05) & 5.1529(+00) & $\rm PA\approx 7\,eV$ & (4)\\
            &        & 6.1187(+00) &-1.6984(-03) & 1.0545(-05) &-1.3659(-08) & 5.6517(-12) & 1.9011(+05) & 5.5956(+00) &&\\
 \ce{ SiN+} & 1340.8 & 1.7963(+00) & 8.7486(-04) &-7.1119(-07) & 2.7522(-10) &-3.2824(-14) & 1.5979(+05) & 1.2276(+01) & $\rm IP\approx 10\,eV$ & (4)\\
            &        & 1.9628(+00) &-2.8399(-03) & 2.3260(-06) & 5.7339(-09) &-5.8556(-12) & 1.6032(+05) & 1.3004(+01) &&\\
 \ce{HNSi+} & 1229.7 & 6.7963(+00) & 8.7486(-04) &-7.1119(-07) & 2.7522(-10) &-3.2824(-14) & 1.4579(+05) &-5.8518(-01) & $\rm PA\approx 7\,eV$ & (4)\\
            &        & 6.9628(+00) &-2.8399(-03) & 2.3260(-06) & 5.7339(-09) &-5.8556(-12) & 1.4632(+05) & 1.4205(-01) &&\\
 \ce{ C3H+} & 1581.9 & 7.3036(+00) & 2.1451(-03) &-1.0729(-06) & 2.6074(-10) &-2.0163(-14) & 1.8901(+05) &-7.5127(-01) & $\rm PA=767.0\,kJ/mol$ & (4)\\
            &        & 7.9329(+00) &-4.4676(-03) & 1.4932(-05) &-1.4795(-08) & 5.0143(-12) & 1.8911(+05) &-2.7279(+00) &&\\
 \ce{C3H2+} & 1416.6 & 6.7541(+00) & 5.2134(-03) &-1.7929(-06) & 2.8083(-10) &-1.6453(-14) & 1.7154(+05) &-8.3646(+00) & $E_{\rm corr}$ & (3)\\
            &        & 3.6062(+00) & 1.8730(-02) &-2.5386(-05) & 1.9449(-08) &-5.9603(-12) & 1.7219(+05) & 6.7567(+00) &&\\
 \ce{  C4+} & 2262.5 & 4.9081(+00) & 3.0190(-03) &-1.1491(-06) & 1.9264(-10) &-1.1833(-14) & 2.7065(+05) &-2.3430(+00) & $\rm IP=12.54\,eV$ & (4)\\
            &        & 1.1833(+00) & 1.7313(-02) &-2.6161(-05) & 2.2430(-08) &-7.8071(-12) & 2.7156(+05) & 1.6135(+01) &&\\
 \ce{ C4H+} & 1806.8 & 9.9081(+00) & 3.0190(-03) &-1.1491(-06) & 1.9264(-10) &-1.1833(-14) & 2.1519(+05) &-1.5204(+01) & $\rm PA=8.03\,eV$ & (3)\\
            &        & 6.1833(+00) & 1.7313(-02) &-2.6161(-05) & 2.2430(-08) &-7.8071(-12) & 2.1611(+05) & 3.2736(+00) &&\\
 \ce{H2NO+} &  951.5 & 5.6660(+00) & 2.9996(-03) &-3.9438(-07) &-3.8534(-11) & 7.0760(-15) & 1.1258(+05) & 6.5045(+00) & $\rm PA=7.17\,eV$ & (3)\\
            &        & 7.0353(+00) &-5.6854(-03) & 1.8520(-05) &-1.7188(-08) & 5.5582(-12) & 1.1243(+05) & 6.0251(-01) &&\\[1mm]
\hline
\end{tabular}}\\[1mm]
\footnotesize
\centering
\begin{minipage}{17cm}
References: $(1)=$ NIST-Janaf \citep{Chase1998}; 
$(2)=$ \citet{Barklem2016};
$(3)=$ \citet{Millar1997};
$(4)=$ NIST (\url{https://webbook.nist.gov/chemistry}).
The NASA seven-polynomials have been constructed or corrected as
\begin{equation}
\vec{F}({\rm A}) \approx \vec{F}({\rm A}) + E_{\rm corr} 
\quad,\quad
\vec{F}({\rm A^+}) \approx \vec{F}({\rm A}) - \vec{F}({\rm el}) + \rm IP \quad,\quad
\vec{F}({\rm AH^+}) \approx \vec{F}({\rm A}) + \vec{F}({\rm H^+}) - \rm PA 
\end{equation}
where $\vec{F}$ is a vector containing the 14 coefficients of the NASA polynomials, $E_{\rm corr}$ an energy correction, IP is the ionisation potential and PA is the proton affinity.  Adding a constant energy means to add the same value [J/mol] to the $6^{\rm th}$ and $12^{\rm th}$ components of $\vec{F}$. 
\end{minipage}
\end{table*}

\section{Disk model parameters}\label{disk_parameters}
We provide the list of input parameters used to model the disk in Table\,\ref{input_parameters}.
\begin{table}[] \label{input_parameters}
\caption{Parameters for the thermochemical disk model used in Sect.\,\ref{Disk modelling}.}
\resizebox{\linewidth}{!}{%
\begin{tabular}{lll} \hline
Quantity                            & Symbol                          & Values                             \\ \hline
\multicolumn{2}{l}{Stellar parameters}                                &                                    \\ \hline
stellar mass                        & $M_{\star}$                     & 0.7\,M$_{\sun}$                   \\
stellar luminosity                  & $L_{\star}$                     & 1\,L$_{\sun}$                   \\
effective temperature               & $T_{\mathrm{eff}}$              & 4000\,K                            \\
UV excess                           & $f_{\mathrm{UV}}$               & 0.01                               \\
UV powerlaw index                   & $p_{\mathrm{UV}}$               & 1.3                                  \\
strength of incident vertical UV    & $\chi^{\mathrm{ISM}}$            & 1                                  \\ 
cosmic-ray \ce{H2} ionization rate  & $\zeta_{\rm CR}$                         & 1.7\,$\times$\,10$^{-17}$ s$^{-1}$ \\ 
X-ray luminosity                     & $L_{\rm{X}}$                & 10$^{30}$\,erg\,s$^{-1}$ \\ \hline
\multicolumn{3}{l}{Disk parameters}                                                                        \\ \hline
minimum dust particle radius        & $a_\mathrm{min}$                & 0.05\,$\mu$m                        \\
maximum dust particle radius        & $a_\mathrm{max}$                & 3000\,$\mu$m                        \\
settling method                     & settle\textunderscore method    & \cite{Riols2018}                  \\
settling parameter                  & $a_\mathrm{settle}$ or $\alpha$ & 10$^{-3}$                          \\
disk gas mass                       & $M_{\mathrm{disk}}$                      & 2.0\,$\times$\,10$^{-2}$\,M$_{\sun}$    \\
dust-to-gas ratio                   & dust-to-gas ratio               & 0.01                               \\
inner disk radius of the outer disk & $R_{\rm in}$                        & 0.05\,au                          \\
outer disk radius                   & $R_{\rm out}$                       & 250\,au                             \\
tapering-off radius                 & $R_{\mathrm{taper}}$            &
30\,au \\
carbon-to-oxygen ratio              & C/O ratio                       & 0.45                          \\
column density power index          & $\epsilon$                      & 1                                  \\
flaring index            & $\beta$                        & 1.15                                  \\
reference scale height    & $H_\mathrm{g}(100~{\rm au})$                           & 10\,au                             \\
extension                           & raduc                           & 1.5                               \\
maximum $\Sigma$ reduction          & reduc                           & 10$^{-6}$                          \\
distance                            & $d$                               & 140\,pc                          \\
inclination                         & $i$                               & 45\textdegree                          \\ 
grid size                           & radial\,$\times$\,vertical            & $200\,\times\,300$                       \\ \hline
\end{tabular}}
\tablefoot{These parameters are explained in detail in \cite{Woitke2009} and \cite{Woitke2023}.} 
\end{table}

\section{Abundances of various species}
\begin{figure*}
   \centering
   \includegraphics[width=0.95\linewidth]{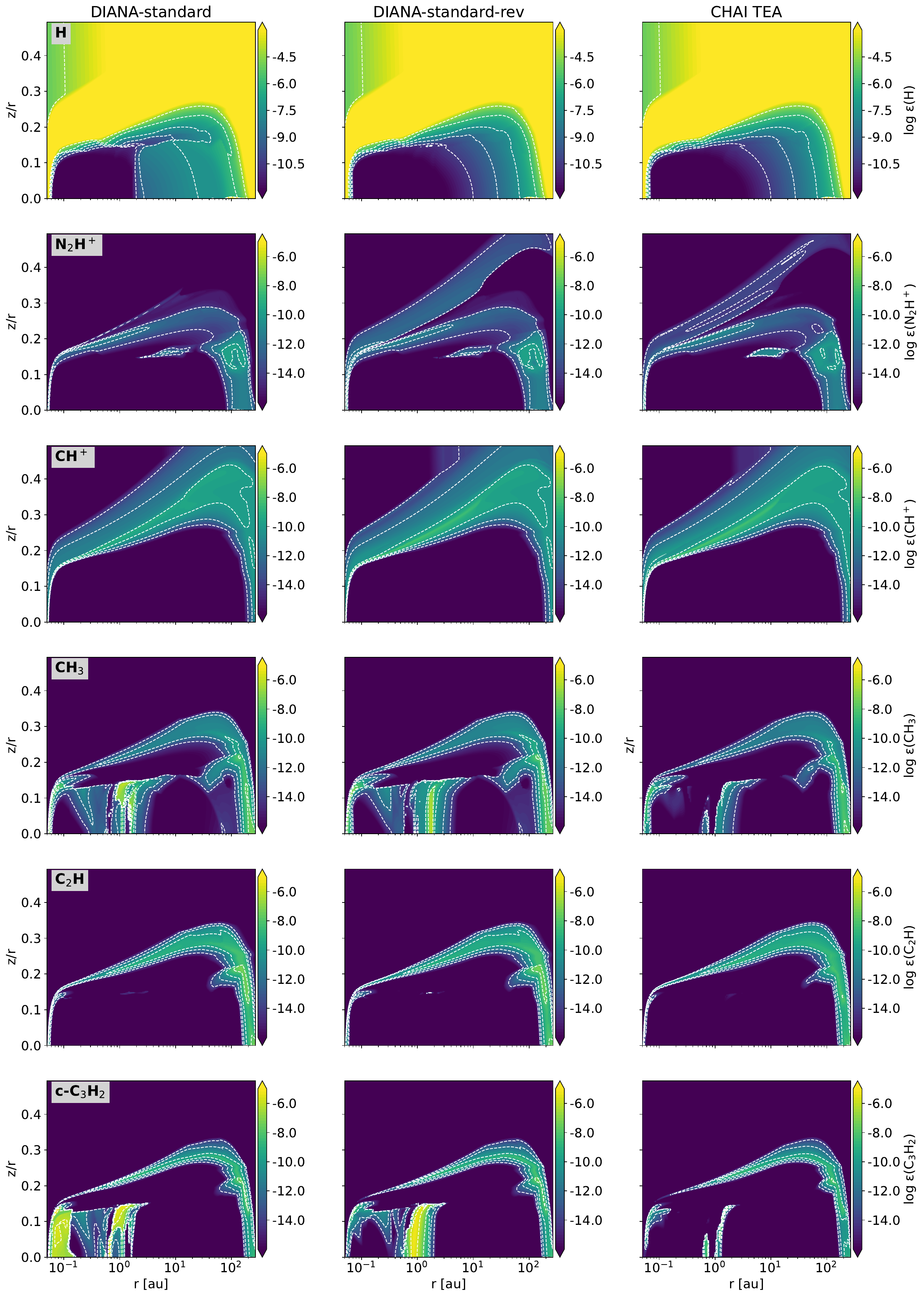}
   \caption{The abundances of various species in the three models calculated using different chemical networks and rate databases outlined in Table\,\ref{tab:rev_models}. The white dashed contours correspond to the tick values on the colourbar.}
  \label{abun_rev_big}
\end{figure*}

Figure\,\ref{abun_rev_big} shows the abundances of various species discussed in Sect.\,\ref{Discussion_ch4}. These abundances are calculated using different chemical networks and rate databases that are outlined in Table\,\ref{tab:rev_models}.




\end{document}